\documentclass{article}
\usepackage[english]{babel}

\usepackage[letterpaper,top=2cm,bottom=2cm,left=3cm,right=3cm,marginparwidth=1.75cm]{geometry}
\usepackage{cite}

\usepackage{amsmath}
\usepackage{amssymb}
\usepackage[dvipdfmx]{graphicx} 
\usepackage{float}
\usepackage{here}
\usepackage{bm}
\usepackage[colorlinks=true, allcolors=blue]{hyperref}
\usepackage[hang,small,bf]{caption}
\usepackage[subrefformat=parens]{subcaption}
\usepackage{color}
\graphicspath{{fig/}}
\usepackage{cleveref}
\usepackage{overpic}

\usepackage{authblk} 

\title{\boldmath The Possibility of Formation of Compact Boson Stars via Cosmological Evolution of a Background Scalar Field}

\author[1]{Yu Miyauchi\thanks{E-mail: miyauchi@tap.scphys.kyoto-u.ac.jp}}
\author[1,2]{Takahiro Tanaka\thanks{E-mail: t.tanaka@tap.scphys.kyoto-u.ac.jp}}

\affil[1]{\small Department of Physics, Kyoto University, Kyoto 606-8502, Japan}
\affil[2]{Center for Gravitational Physics and Quantum Information, Yukawa Institute for Theoretical Physics, Kyoto University, Kyoto 606-8502, Japan}
\date{}%

\begin{document}

\maketitle

\begin{abstract}
     Boson stars, hypothetical astrophysical objects bound by the self-gravity of a scalar field, have been widely studied as a type of exotic compact object that is horizonless and provides a testing ground for physics beyond the Standard Model.
     In particular, many previous works have demonstrated methods for distinguishing compact boson stars from black holes in general relativity through gravitational wave observations.
     However, the formation scenario of compact boson stars within the age of the universe remains unclear. 
     In this paper, we explore a possible scenario for the formation of compact boson stars. 
     The model we consider requires two coupled scalar fields: a complex scalar field that forms a boson star and a spatially homogeneous background field, as formation of a compact boson star cannot be achieved in a single filed model.
     Using the adiabatic approximation, we show that non-relativistic boson clouds can evolve into compact boson stars through the cosmological time-evolution of the background field. 
     In our model the background field evolves to increase the effective mass of the scalar field, and as a result compact boson stars can form within the cosmological timescale, if the variation of the background field is as large as the Planck scale.
    However, further investigation is required because the required initial states are not the configurations that can be described by the well-studied Schr\"odinger-Poisson system.
\end{abstract}

\section{Introduction}
\label{sec_intro}
\par
Boson stars are objects composed of a scalar field bound by self-gravity.
For example, mini-boson star solutions \cite{PhysRev.172.1331} can be constructed by a free massive complex scalar field described by the Lagrangian 
\begin{equation}
\label{eq_mini}
\mathcal{L} = -\frac{1}{2}|\partial_\mu \Phi|^2 - \frac{1}{2}m_\phi^2|\Phi|^2.
\end{equation}
Here, $m_\phi$ represents the mass of the $\Phi$-field.
Various objects composed of scalar fields have been proposed, including boson stars with a $|\Phi|^4$-term \cite{PhysRevLett.57.2485}, solitons formed by a real scalar field such as oscillatons \cite{Seidel:1991zh}, and non-topological soliton stars \cite{Kunz:2021mbm, Endo:2022uhe, Tamaki:2011zza, Collodel_2022}, which is a non-gravitationally localized configurations of a scalar field.
Non-topological solitons include Q-balls \cite{Coleman:1985ki} and the Friedberg–Lee–Sirlin soliton model, which is composed of two interacting scalar fields \cite{Friedberg:1976me}.
\par
Boson stars composed of a complex scalar field can form coherent configurations that carry a conserved global $U(1)$ charge. 
The stability of such configurations is supported by a large conserved particle number. 
To realize such stable configurations, a charge asymmetry between particles and antiparticles is essential.
In many studies of complex scalar boson stars, the presence of this charge asymmetry is taken as a given.
One well-known mechanism for generating such asymmetries is the Affleck–Dine mechanism, originally proposed in the context of baryogenesis~\cite{Affleck:1984fy}.
Such asymmetries can also lead to the formation of Q-balls~\cite{Kasuya:1999wu}.
For simplicity, in this work we focus on boson stars composed of a complex scalar field.

\par
Axions, well-known candidates for dark matter, have been proposed as scalar fields that form  boson stars. 
Axions are expected to distribute across a wide mass spectrum, $10^{-3}-10^{-33}\,\mathrm{eV}$, and are thought to leave various signatures depending on their mass scale.
This paradigm is referred to as the axiverse \cite{Arvanitaki:2009fg}. 
Scalar fields with masses of the same order as the current Hubble constant, $H_0 \simeq 10^{-33}\,\mathrm{eV}$, are candidates for dark energy \cite{Choi:2021aze}.
QCD axions with the mass around $10^{-6}\,\mathrm{eV}$ are expected to develop gravitationally bound systems, known as axion miniclusters, during the radiation-dominated era and boson stars can subsequently form within the minicluster \cite{Hogan:1988mp, Kolb:1993zz}.
During the matter-dominated era, these miniclusters can merge to form larger dark matter halos \cite{PhysRevD.107.083510}.
Similarly, Fuzzy dark matter, characterized by the mass ranging $m_\phi=10^{-22}-10^{-20}~\mathrm{eV}$, is a compelling candidate for explaining the formation of galactic soliton cores within dark matter halos \cite{Hui:2021tkt}.
Numerical simulations have demonstrated that such non-relativistic boson clouds naturally emerge under various conditions \cite{Schive:2014dra,PhysRevLett.121.151301,Chan:2022bkz,Eggemeier:2019jsu}. 
Their formation mechanism is attributed to gravitational scattering induced by density fluctuations in the scalar field, a process known as gravitational cooling~\cite{Seidel:1993zk,Guzman:2006yc}.
In this mechanism, the initial configuration loses kinetic energy through scalar field radiation, resulting in the formation of a virialized scalar configuration.
The timescale for the formation of these clouds can be estimated using the two-body relaxation time associated with gravitational scattering.
For instance, the relaxation time in the fuzzy dark matter model is given by
\begin{equation}
\label{eq_relaxationtime}
    t_\mathrm{rel} 
     =\frac{\sqrt{2}b}{12\pi^3\log(m_\phi v L)}\frac{m_\mathrm{pl}^4m_\phi^3v^6}{\rho_g^2}
    \simeq 10^6\,\mathrm{yr} \left( \frac{m_\phi}{10^{-22}\,\mathrm{eV}} \right)^3 \left( \frac{v}{30\,\mathrm{km/s}} \right)^6 \left( \frac{0.1\,\mathrm{M_\odot}/\mathrm{pc}^3}{\rho_g} \right)^2,
\end{equation}
where $v$ is the velocity dispersion, $\rho_g$ is the energy density, $b=\mathcal{O}(1)$ is a numerical factor that depends on the initial conditions, and $L$ is the box size~\cite{PhysRevLett.121.151301,Chan:2022bkz}.
The formation timescale of non-relativistic boson clouds within axion miniclusters also follows the middle expression of Eq.~\eqref{eq_relaxationtime}.
Taking into account the axion minicluster formation scenario described in Ref.~\cite{Kolb:1993zz}, it is given by
\begin{equation}
\label{eq_relaxationtime2}
    t_\mathrm{rel} \simeq \frac{10^9\,\mathrm{yr}}{\Phi^3(1+\Phi)} \left( \frac{m_\phi}{26\,\mathrm{\mu eV}} \right)^3 \left( \frac{M_c}{10^{-13}\,\mathrm{M_\odot}} \right)^2 ,
\end{equation}
where $M_c$ denotes the mass of an axion minicluster and $1+\Phi$ is the ratio of its central density to the background density in the radiation-dominated era~\cite{PhysRevLett.121.151301}.
Consequently, the formation and properties of non-relativistic boson clouds that constitute dark matter have been the focus of extensive research.
Such formation mechanisms are also applicable to complex scalar fields, since their self-gravitating configurations reduce to the Schrödinger–Poisson system in the non-relativistic limit, as in the case of axions.
\par
Boson stars are known to exist as stable compact objects that resist gravitational collapse \cite{PhysRevD.42.384, Hawley:2000dt, Gleiser:1988ih}.
If boson stars become compact objects, they could serve as horizonless compact objects.
This prospect has motivated extensive studies to identify potential differences between compact boson stars and black holes in general relativity through gravitational wave observations (see review in Ref.~\cite{Cardoso:2019rvt}).
For example, the tidal Love numbers of binary compact boson stars were calculated in Ref.~\cite{Cardoso:2017cfl}.
The coalescence of binary compact boson stars including the resulting gravitational wave signals \cite{Palenzuela:2017kcg} and quasi-normal modes of compact boson stars in comparison to those of black holes \cite{Macedo:2013jja} were also investigated.
In particular, compact boson stars composed of scalar fields with $m_\phi = \mathcal{O}(10^{-10}\,\mathrm{eV})$ have a total mass, $M = \mathcal{O}(1\,\mathrm{M_\odot})$. 
This mass range falls within the detectable regime of current gravitational wave observatories, such as the LIGO-Virgo-KAGRA (LVK) collaboration.
Future gravitational wave observatories, such as the space-based detectors LISA, TianQin, Taiji, and B-DECIGO, as well as the next-generation ground-based detectors Einstein Telescope (ET) and Cosmic Explorer (CE), are anticipated to play a crucial role in investigating a wider mass range.
They will constrain the parameters of scalar fields as dark matter candidates, opening new avenues for understanding the universe \cite{Giudice:2016zpa}.

\par
However, the formation scenario of compact boson stars remains unclear, as explained in more detail in the following section. 
First, we will see that the gravitational cooling mechanism cannot directly lead to the condensation of scalar field fluctuations into stellar mass compact boson stars. 
Therefore, it is necessary to consider an alternative possibility for compact boson stars to evolve from non-relativistic boson clouds.
For example, as a growth mechanism for boson clouds, gravitational scattering between density fluctuations of unbounded scalar fields and the scalar field constituting the cloud has been proposed, similar to the formation mechanisms discussed earlier. 
However, it has been shown that the timescale for such growth is significantly longer than the formation timescale \eqref{eq_relaxationtime}
, exceeding the age of the universe.
As we shall see below, growth due to self-interactions is also too slow to form compact boson stars within cosmological timescales.

\par  
In this paper, we explore a possible scenario in which non-relativistic boson clouds can evolve into compact boson stars within the cosmological timescale. 
Previous works on boson star formation have typically focused on single scalar field models. 
The main challenge in such scenarios is that the growth of boson stars through gravitational or non-gravitational interactions is generally too slow to produce compact configurations within the age of the Universe.
However, the existence of multiple scalar fields is a common feature in theories beyond the Standard Model (e.g., the axiverse), motivating to introduce couplings between scalar fields with hierarchical masses. 
As a simple model, we consider a system in which a complex scalar field, $\phi$ constituting boson stars, is coupled to a real scalar field, $\chi$, representing a cosmologically evolving background field.
The $\chi$-field should have a sufficiently small mass to undergo slow-roll motion on an appropriate potential $V_\mathrm{SR}$. 
We will show that the slow-roll evolution of the $\chi$-field, through interactions between the scalar fields, can render the boson star more compact.
We also quantify the amount of change in the $\chi$-field required for the boson star to become compact enough.
Based on the adiabatic approximation, we assume that the slow-roll variation of the $\chi$-field corresponds to a change in the outer boundary condition of the local $\chi$-field configuration around the boson star.
We construct a series of stationary solutions for different boundary values of the $\chi$-field.
Using these solutions, we trace adiabatic evolutionary paths to investigate whether boson stars can attain a more compact configuration.
For this purpose, the initial and final states of the adiabatic evolution are to be set to the threshold for the evaporation of a non-relativistic boson cloud, and that of gravitational instability, respectively. 
As a result, we will find that if the $\chi$-field undergoes slow-roll motion such that increases the effective mass of the $\phi$-field, the boson star becomes more compact. 
We will find that non-relativistic boson clouds can evolve into compact boson stars, if the variation of the $\chi$-field is as large as $\mathcal{O}(m_\mathrm{pl})$.
However, we will also find that these initial states are significantly different from the fiducial configurations that are well-approximated by the Schr\"odinger-Poisson system.

\par
The plan of the paper is as follows.
In Sec.~\ref{sec_Difficulty in the Formation of Compact Boson Stars Via Gravitational Scattering}, we discuss the difficulty of forming compact boson stars either through the direct collapse of scalar field density fluctuations or through mass growth via two-body scattering. 
In the latter case, we show that it is not possible for compact boson stars to form via gravitational scattering within the age of the Universe, based on the results presented in Ref.~\cite{Chan:2022bkz} and the consideration of $\phi^4$ self-interactions.
In Sec.~\ref{sec_Analysis of Stationary Configurations}, we derive the basic equations and boundary conditions that govern the $\phi$-and $\chi$-fields mentioned above, and the two spacetime metric components, which are essential in obtaining the stationary solutions of the boson stars.
In Sec.~\ref{sec_Initial State and Final State of Adiabatic Evolution}, we discuss the selection of initial and final states for the adiabatic evolution of the boson stars. 
Sec.~\ref{sec_results} is divided into three subsections. 
In Sec.~\ref{sec_Properties of the sequences of the stationary configurations for each chiout}, we investigate the differences in the properties of the configurations by analyzing the sequences of stationary solutions obtained numerically for various values of the $\chi$-field boundary condition.
In Sec.~\ref{sec_Adiabatic Evolution Paths and Mechanisms}, we explore the adiabatic evolution paths and demonstrate that the boson stars can become more compact through this process. 
We also find regions where the gradient of the $\chi$-field cannot be ignored in the initial conditions.
In Sec.~\ref{sec_Evaluation of the Change in Adiabatic Evolution}, we evaluate the required change in the $\chi$-field, and  
analyze how the effective mass of the $\phi$-field and the total mass $M$ of the boson star change throughout this evolutionary process.
In Sec.~\ref{sec_Discussion and Conclusion}, we summarize our results and discuss the form of slow roll potentials $V_\mathrm{SR}$ that can make boson stars compact enough within the age of the universe. 
We also note that further investigation is required because the required initial states are not the configurations that can be described by the well-studied Schrödinger–Poisson system.
In this paper, we use natural units with $c = \hbar = 1$.

\section{Difficulty in the Formation of Compact Boson Stars in the Universe}
\label{sec_Difficulty in the Formation of Compact Boson Stars Via Gravitational Scattering}

As mentioned in Sec.~\ref{sec_intro}, it is generally considered difficult for a compact boson star to form in the Universe.
We begin by showing that such objects are unlikely to form via direct gravitational collapse.
Assuming that gravitational attraction is the dominant binding force, the gradient energy of the $\phi$-field is balanced by its gravitational potential energy:
\begin{equation}
\label{eq_dekinai1}
    \frac{\phi_0^2}{R^2}R^3\simeq \frac{M^2}{m_\mathrm{pl}^2R},
\end{equation}
where $\phi_0$, $M$ and $R$ are the central value of the $\phi$-field, the total mass, and the radius of the configuration, respectively. Here, we assume that the asymptotic value of the $\phi$-field vanishes. 
Using the definitions of the compactness, $C = M/(m_\mathrm{pl}^2 R)$, we obtain
\begin{equation}
\label{eq_dekinai_comapct}
    C \simeq \frac{\phi_0}{m_\mathrm{pl}}.
\end{equation}
Thus, to achieve $C = \mathcal{O}(1)$, we require
\begin{equation}
\label{eq_dekinai_mpl}
    \phi_0 \simeq m_\mathrm{pl}.
\end{equation}
On the other hand, for the $\phi$-field to begin oscillating and behave as dark matter, its temporal change rate $\dot{\phi}/\phi$ needs to be as large as the Hubble parameter $H$.
Using the slow-roll equation and the Friedmann equation,
\begin{equation}
    3H\dot{\phi} \simeq \frac{dV(\phi)}{d\phi},\quad H^2 = \frac{8\pi}{3m_\mathrm{pl}^2}\rho_\mathrm{tot},
\end{equation}
this condition can be rewritten as
\begin{equation}
\label{eq_dekinai_osci_condition}
    \frac{\dot{\phi}}{H\phi} \simeq \frac{dV(\phi)}{H^2 \phi d\phi} \simeq \left(\frac{m_\mathrm{pl}}{\phi}\right)^2 \frac{\rho}{\rho_\mathrm{tot}} \simeq 1\,,
\end{equation}
where $\rho$ is the energy density of the $\phi$-field. 
We consider that the $\phi$-field starts to move during the radiation-dominated era, in which the energy density must satisfy $\rho \ll \rho_r \simeq \rho_\mathrm{tot}$, where $\rho_r$ is the energy density of radiation.
Hence, Eq.~\eqref{eq_dekinai_osci_condition} implies
\begin{equation}
\label{eq_dekinai_vev}
    \phi \ll m_\mathrm{pl}.
\end{equation}
From Eq.~\eqref{eq_dekinai_mpl} and Eq.~\eqref{eq_dekinai_vev}, we conclude that compact boson stars are quite unlikely to form directly from cosmological initial conditions.

\par
\par
For example, if the $\phi$-field corresponds to an axion, its density fluctuations $\delta \rho$ are isocurvature, and therefore do not grow during the radiation-dominated era.
However, on scales much smaller than the Hubble horizon, $\delta\rho$ can locally exceed $\rho_r$, since the background radiation dilutes. 
When $\delta\rho \gtrsim \rho_r$ is satisfied, gravitational collapse can occur within such local over-density regions, leading to the formation of axion miniclusters.
During the virialization process, the size of an axion minicluster typically becomes about half the size of the over-density region. 
As a result, the order of magnitude of complactness $C$ remains unchanged before and after collapse.
Consequently, the amplitude of the $\phi$-field within axion miniclusters must still satisfy $\phi \ll m_\mathrm{pl}$.
According to Eq.~\eqref{eq_dekinai_mpl}, this implies that compact boson stars cannot form directly from such miniclusters via two-body gravitational relaxation; instead, only non-relativistic boson clouds can form through this process.

\par
Next, we point out that the growth of non-relativistic boson clouds through gravitational scattering is too slow to form compact boson stars within the age of the Universe.
Many previous numerical studies have demonstrated that non-relativistic boson clouds can spontaneously form on the relaxation timescale through gravitational scattering with unbound scalar particles, {\it i.e.}, boson gas.
Therefore, a natural scenario of the formation of compact boson stars is that a non-relativistic boson cloud grows in mass via gravitational scattering with the surrounding boson gas.
In Ref.~\cite{Chan:2022bkz}, the authors analyzed this process in the Schr\"odinger–Poisson system and evaluated the mass change rate of the boson cloud with total mass $M$,
\begin{equation}
\label{eq_gamma_def}
    \Gamma = \frac{1}{M} \frac{dM}{dt}.
\end{equation}

\par
In Ref.~\cite{Chan:2022bkz}, a dimensionless parameter
\begin{equation}
\label{eq_nu_def}
\nu = \frac{k_g}{k_s}\,,
\end{equation}
was introduced, where $k_g$ is the typical momentum of the boson gas defined by
\begin{equation}
k_g = m_\phi v,
\end{equation}
$m_\phi$ is the effective mass of the scalar field, and $v$ is the typical velocity dispersion of a galactic halo. 
Following Ref.~\cite{Chan:2022bkz}, we assume $v = 10^{-3}$ in this paper.
$k_s$ is the typical momentum of the boson cloud.
It was shown that the boson cloud evaporates ($\Gamma<0$) for $\nu \gg 1$, while it grows ($\Gamma>0$) for $\nu \ll 1$.

The Schrödinger-Poisson system is invariant under a certain scaling transformation \cite{Marsh:2015wka}. 
Based on this invariance, Ref.~\cite{Chan:2022bkz} estimated the mass of a boson cloud as
\begin{equation}
\label{eq_ks_def}
M = \mu_0 \frac{k_s m_\mathrm{pl}^2}{4 \pi m_\phi^2},
\end{equation}
where $\mu_0 = 25.9$ is a numerical constant.
Eq.~\eqref{eq_nu_def} implies that $\nu \gtrsim 1$ corresponds to light boson clouds, while $\nu \ll 1$ corresponds to heavy boson clouds.
For $\nu \ll 1$, the growth rate of a heavy boson cloud is given by ~\cite{Chan:2022bkz}
\begin{equation}
\label{eq_growth_rate}
    \Gamma = A \frac{(4\pi)^2 m_\phi^3 \rho_g^2}{m_\mathrm{pl}^4 k_g^6} \nu^4,
\end{equation}
where $\rho_g$ is the typical energy density of the boson gas, and $A(>0)$ is a numerical coefficient of $O(1)$.
\par
For the formation of compact boson stars, the timescale associated with Eq.~\eqref{eq_growth_rate} must be shorter than the Hubble time $H_0^{-1}$, which can be expressed as
\begin{equation}
\label{eq_growth_h0}
    \frac{\Gamma}{H_0} \gtrsim 1.
\end{equation}
To rewrite Eq.~\eqref{eq_growth_h0}, we use the Friedmann equation,
\begin{equation}
\label{eq_fried_current}
    H_0^2 = \frac{8\pi}{3m_\mathrm{pl}^2} \rho_\mathrm{crit,0},
\end{equation}
where $\rho_\mathrm{crit,0}$ is the current critical density.
In addition, for $C = \mathcal{O}(1)$, the typical size of the object is comparable to the Compton wavelength of the scalar field, implying $k_s \simeq m_\phi$.
Substituting these relations into Eq.~\eqref{eq_growth_h0}, and ignoring the unknown numerical coefficient $A$, we obtain
\begin{equation}
\label{eq_growth_h02}
    \frac{\Gamma}{H_0} \simeq \left( \frac{\Omega_\mathrm{DM,0}}{v} \right)^2 \left( \frac{H_0}{m_\phi} \right)^3 \gtrsim 1,
\end{equation}
where $\Omega_\mathrm{DM,0} \simeq \rho_g / \rho_\mathrm{crit,0} ~(= \mathcal{O}(0.1))$ is the current dark matter density parameter.
To satisfy this condition, the scalar field mass must be
\begin{equation}
\label{eq_growth_h03}
    m_\phi \lesssim H_0\left(\frac{\Omega_{\mathrm{DM,0}}}{v}\right)^{\frac{2}{3}} \simeq10^{-33}~\mathrm{eV}\left(\frac{\Omega_{\mathrm{DM,0}}}{v}\right)^{\frac{2}{3}}.
\end{equation}
$v$ represents the virial velocity associated with the structure under consideration. 
In this study, we adopt $v \simeq 10^{-3}$, corresponding to large galaxies. 
In smaller structures, $v$ becomes significantly lower; for instance, axion miniclusters typically have $v \simeq 10^{-7}$~\cite{PhysRevLett.121.151301}.
In all such cases, the value of $m_\phi$ required to satisfy Eq.~\eqref{eq_growth_h03} is far outside the range of stellar mass boson stars.
Therefore, we conclude that a compact boson star cannot form within a Hubble time through gravitational scattering. 
\par
In addition to Ref.~\cite{Chan:2022bkz}, several numerical simulations have investigated the growth of non-relativistic boson clouds \cite{Chen:2020cef, Dmitriev:2023ipv, Liao:2024zkj, Blum:2025aaa}.
These studies consistently reproduce the formation timescale and initial growth rate reported in Ref.~\cite{PhysRevLett.121.151301}. 
However, the late-time behavior of boson clouds varies across the literature. Ref.~\cite{Chan:2022bkz} derives a power-law growth of the mass $M$ from Eq.~\eqref{eq_growth_rate}, using Eqs.~\eqref{eq_gamma_def} and~\eqref{eq_ks_def}:
\begin{equation}
 M \propto t^n, \quad n = \frac{1}{4}.
\end{equation}
In contrast, Refs.~\cite{Chen:2020cef} and~\cite{Dmitriev:2023ipv} report shallower growth rates, with $n = 1/8$ and $n = 1/9$, respectively. 
Furthermore, Refs.~\cite{Liao:2024zkj, Blum:2025aaa} show that the growth eventually saturates at late times.
These results suggest that this mechanism is insufficient for boson clouds to evolve into compact boson stars within the age of the Universe.
\par
So far, we have considered growth scenarios driven by the gravitational scattering. 
We now show that compact boson stars cannot form even when self-interactions of the $\phi$-field are included. 
As a simple example, we consider an axion-like cosine potential:
\begin{equation}
    V(\phi) = m_\phi^2 f^2 \left(1 - \cos\frac{\phi}{f}\right).
\end{equation}
Nonlinear effects become important when $\phi/f \simeq 1$. 
Applying Eq.~\eqref{eq_dekinai_comapct}, the compactness is given by
\begin{equation}
    C \simeq \frac{f}{m_\mathrm{pl}}.
\end{equation}
Therefore, achieving $C = \mathcal{O}(1)$ requires
\begin{equation}
\label{eq_dekinai_self1}
    f \simeq m_\mathrm{pl}.
\end{equation}
Next, we estimate the growth rate due to $\phi^4$ self-interactions, $\Gamma_\mathrm{self}$. 
The cross sections for gravitational and self-interaction scatterings are, respectively, given by
\begin{equation}
    \sigma_\mathrm{gr} \simeq \frac{m_\phi^2}{m_\mathrm{pl}^4 v^4}, \quad \sigma_\mathrm{self} \simeq \frac{m_\phi^2}{f^4},
\end{equation}
where the Coulomb logarithm for gravitational scattering is neglected. 
Simply replacing $\sigma_\mathrm{gr}$ with $\sigma_\mathrm{self}$, 
the growth rate due to self-interaction can be estimated using Eq.~\eqref{eq_relaxationtime} as
\begin{equation}
\label{eq_self_rate}
    \Gamma_\mathrm{self} \simeq \frac{\sigma_\mathrm{self}}{\sigma_\mathrm{gr}} \cdot \frac{1}{t_\mathrm{rel}} \simeq \frac{\rho_g^2}{m_\phi^3 v^2 f^4}.
\end{equation}
Similarly to Eq.~\eqref{eq_growth_h0}, the condition 
\begin{equation}
    \frac{\Gamma_\mathrm{self}}{H_0} \gtrsim 1,
\end{equation}
must be satisfied.
Substituting Eq.~\eqref{eq_self_rate} and using Eq.~\eqref{eq_fried_current}, we obtain
\begin{equation}
    \left( \frac{\Omega_\mathrm{DM,0}}{v} \right)^2 \left( \frac{H_0}{m_\phi} \right)^3 \gtrsim \left( \frac{f}{m_\mathrm{pl}} \right)^4.
\end{equation}
For the mass range of the $\phi$-field under consideration, this inequality implies
\begin{equation}
\label{eq_dekinai_self2}
    f \ll m_\mathrm{pl}.
\end{equation}
Combining Eqs.~\eqref{eq_dekinai_self1} and \eqref{eq_dekinai_self2}, we conclude that the formation of compact boson stars is not possible even with self-interaction-driven growth.
\par
A related line of research has recently investigated the possibility that bosonic dark matter captured by neutron stars could lead to the formation of solar-mass black holes~\cite{Garani:2021gvc}. 
Ref.~\cite{Garani:2021gvc} proposed a mechanism in which the captured dark matter undergoes gravitational collapse and forms a black hole. 
As in our scenario of compact boson star formation, their approach seeks to resolve the tension between the energy scales required for gravitational collapse and those associated with dark matter self-interactions.
In single-field scalar models, these requirements are generally incompatible: gravitational collapse favors extremely weak self-interactions (cf. Eq.~\eqref{eq_dekinai_self1}), whereas efficient capture and thermalization within neutron stars require relatively strong self-interactions (cf. Eq.~\eqref{eq_dekinai_self2}).
To overcome this conflict, both Ref.~\cite{Garani:2021gvc} and our work introduce two scalar fields, but through different mechanisms.
In Ref.~\cite{Garani:2021gvc}, the self-interacting scalar field $\phi$ accumulates inside the neutron star and undergoes a potential-induced phase transition into an auxiliary field $\chi$ with suppressed self-interactions. 
In contrast, our model employs the cosmological evolution of the $\chi$ - field to dynamically resolve the tension between the competing scales, as will be detailed in the following sections.

\section{Analysis of Stationary Configurations}
\label{sec_Analysis of Stationary Configurations}
\begin{figure}[t]
\centering
\includegraphics[width=0.6\linewidth]{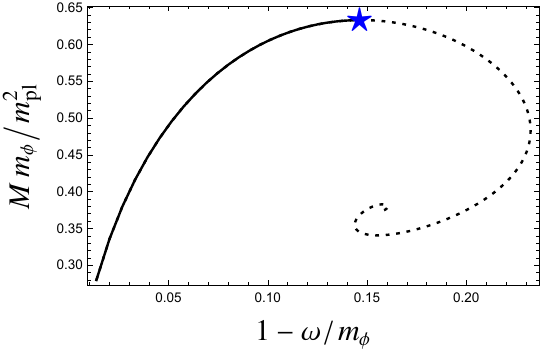}
\caption{
The total mass $M$ of stationary solutions for mini-boson stars where the Lagrangian is given by Eq.~\eqref{eq_mini}.
The horizontal axis represents $1-\omega/m_\phi$, where $m_\phi$ is the mass of the $\phi$-field and $\omega$ is the eigenvalue defined by Eq.~\eqref{eq_scalar}. 
In mini-boson star cases, the stationary configurations form a one-parameter family, parameterized by either the central field value $\phi_0=\phi(0)$ or the eigenvalue $\omega$. 
The blue point at $\omega/m_\phi=0.85$ represents the maximum mass of mini-boson stars as well as the critical point for gravitational instability.
The solid curve represents the stable branch, while the dashed curve represents the unstable branch. 
}
\label{fig_mo1plot}
\end{figure}

We consider a self-gravitating system which consists of a complex scalar field $\Phi$ and a background real scalar field $\chi$.
To trace the adiabatic evolution of a boson star, we numerically solve stationary configurations.
The Lagrangian is give by
\begin{equation}
\label{eq_lagrangian}
\mathcal{L} = -\frac{1}{2}|\partial_\mu\Phi|^2 - \frac{1}{2}(\partial_\mu\chi)^2 - V(\Phi, \chi),
\end{equation}
with the interaction potential 
\begin{equation}
\label{potential}
V(\Phi, \chi) = \frac{1}{2}g^2\chi^2|\Phi|^2.
\end{equation}
The asymptotic value of the $\chi$-field is assumed to roll down an appropriate potential $V_\mathrm{SR}(\chi)$ slowly.
The effects of the potential $V_\mathrm{SR}(\chi)$ are considered negligible in the vicinity of the boson star and are therefore omitted in the Lagrangian \eqref{eq_lagrangian}.
The adiabatic time-evolution of the $\chi$-field outside the boson star is taken into account by varying the outer boundary condition for $\chi$-field.
We assume a stationary and static configuration,
\begin{equation}
\label{eq_scalar}
\Phi(t, r) = \phi(r)e^{-i\omega t}, \quad \chi(t, r) = \chi(r).
\end{equation}
The effective mass of the $\phi$-field, $m_\phi$, is expressed as
\begin{equation}
\label{mphi}
m_\phi(r) = g\chi(r),
\end{equation}
which varies spatially.
If the spatial gradient of $\chi$-field is zero, the Lagrangian~\eqref{eq_lagrangian} coincides with that of mini-boson stars.
We denote  the static, spherically symmetric spacetime metric as 
\begin{equation}
\label{eq_metric}
ds^2 = -N^2(r)dt^2 + G^2(r)dr^2 + r^2d\Omega.
\end{equation}
\par
Since the Lagrangian \eqref{eq_lagrangian} is invariant under global $U(1)$ gauge transformations, the current
\begin{equation*}
    J^\mu=\frac{\partial\mathcal{L}}{\partial(\partial_\mu\Phi)}\delta\Phi+\frac{\partial\mathcal{L}}{\partial(\partial_\mu\Phi^*)}\delta\Phi^*
    =\frac{i}{2}g^{\mu\nu}\left(\Phi^*\partial_\nu\Phi-\Phi\partial_\nu\Phi^*\right)
\end{equation*}
satisfies the conservation law $\bm{\nabla}_\mu J^\mu = 0$. 
The conserved charge is then
\begin{equation}
\label{eq_particle number}
J = \int J^0\sqrt{-g}dx^3 = 4\pi\omega\int\frac{G(r)}{N(r)}\phi^2(r)r^2dr,
\end{equation}
which represents the total particle number of the $\phi$-field.
$J$ is also an adiabatic invariant, which remains constant throughout the adiabatic evolution.
\par
From the $(0,0)$ and $(1,1)$ components of the Einstein equations $G_{\mu\nu}=8\pi/m^2_\mathrm{pl}T_{\mu\nu}$, we derive the equations governing $N(r)$ and $G(r)$. The components of the Einstein tensor are
\begin{align}
    &G_{00}=\frac{N^2}{rG^4}\left[\frac{G^2(G^2-1)}
{r}+(G^2)'\right],&
\\
&G_{11}=\frac{1}{rG^2}\left[-\frac{N^2(G^2-1)}
{r}+(N^2)'\right],&
\end{align}
where $'$ denotes the derivative with respect to $r$.
The energy-momentum tensor is expressed as
\begin{align*}
    T_{\mu\nu}=&\frac{1}{2}\left(\partial_\mu\Phi\partial_\nu\Phi^*+\partial_\nu\Phi\partial_\mu\Phi^*\right)+\partial_\nu\chi\partial_\mu\chi&
    \\
    &-\frac{1}{2}g_{\mu\nu}\left(\partial_\alpha\Phi\partial^\alpha\Phi^*+\partial_\alpha\chi\partial^\alpha\chi+2V\right),&
\end{align*}
and its $(0,0)$ and $(1,1)$ components are
\begin{align}
\label{eq_t00}
   & T_{00}=\frac{1}{2}\omega^2\phi^2+\frac{1}{2}\frac{N^2}{G^2}\Bigl(\phi'^2+\chi'^2\Bigr)+N^2V,&
   \\
    & T_{11}=\frac{1}{2}\frac{G^2}{N^2}\omega^2\phi^2+\frac{1}{2}\bigl(\phi'^2+\chi'^2\bigr)-G^2V.&
\end{align}
Thus, the Einstein equations yield the following differential equations for the metric functions $G(r)$ and $N(r)$:
\begin{align}
\label{eq_eom_g}
    &G'=\frac{G}{2}\left[\frac{(1-G^2)}{r}+\frac{4\pi r}{m_\mathrm{pl}^2} \left(G^2\bigl(2V+\frac{\omega^2}{N^2}\phi^2\bigr)+\phi'^2+\chi'^2\right)\right],&
    \\
\label{eq_eom_n}
     &N'=\frac{N}{2}\left[\frac{(G^2-1)}{r}+\frac{4\pi r}{m_\mathrm{pl}^2}\left(G^2\left(\frac{\omega^2}{N^2}\phi^2-2V\right)+\phi'^2+\chi'^2\right)\right].&
\end{align}
The Klein-Gordon equations governing $\phi(r)$ and $\chi(r)$ are
\begin{align}
   \label{eq_eom_phi}
   N^2 \phi'' + N^2 \left( 1 + G^2 - 2\frac{4\pi}{m_\mathrm{pl}^2} r^2 G^2  V \right) \frac{\phi'}{r} + G^2\left( \omega^2 \phi - N^2 \frac{dV}{d\phi} \right) =0,
   \\
   \label{eq_eom_chi}
   N^2 \chi'' + N^2 \left( 1 + G^2 - 2\frac{4\pi}{m_\mathrm{pl}^2}r^2 G^2  V \right) \frac{\chi'}{r}  -N^2G^2\frac{dV}{d\chi}=0.
\end{align}
\par
To obtain the stationary solutions for the boson star in this model, appropriate boundary conditions must be imposed. 
At the center $(r=0)$, regularity imposes
\begin{equation}
\label{eq_bc_ori}
    \phi'(0)=0,\quad \chi'(0)=0,\quad G(0)=1.
\end{equation}
At spatial infinity $(r = \infty)$, asymptotic flatness imposes
\begin{equation}
\label{eq_bc_inf}
    \phi(\infty)=0,\quad N(\infty)=G(\infty)=1,\quad     \chi(\infty)=\chi_\mathrm{out}.
\end{equation}
Here, $\chi_\mathrm{out}$ is the background value of the $\chi$-field outside the boson star at a given moment of time. 
In mini-boson star cases, the stationary configurations form a one-parameter family, parameterized by either the central field value $\phi_0 = \phi(0)$ or the eigenvalue $\omega$ (Fig.~\ref{fig_mo1plot}). 
In our model \eqref{eq_lagrangian}, the stationary configurations depend on two parameters: $\phi_0$ (or $\omega$) and $\chi_\mathrm{out}$.

\par
By using  the quantity with mass dimension, such as $m_\mathrm{pl}/\sqrt{4\pi}$, we introduce the dimensionless variables by
\begin{equation}
\label{eq_scaling}
    \omega= \frac{gm_\mathrm{pl}\tilde{\omega}}{\sqrt{4\pi}},\quad 
    r = \frac{\sqrt{4\pi}\tilde{r}}{gm_\mathrm{pl}}, \quad \phi^2=\frac{m_\mathrm{pl}^2\tilde{\phi}^2}{4\pi}, \quad \chi^2= \frac{m_\mathrm{pl}^2 \tilde{\chi}^2}{4\pi}.
\end{equation}
Using the rescaled potential defined by
\begin{equation}
    V= \frac{g^2m_\mathrm{pl}^4\tilde{V}}{(4\pi)^2},
\end{equation}
Eqs. \eqref{eq_eom_g}-\eqref{eq_eom_chi} become
\begin{align}
\label{eq_eom_g2}
 &G_{,\tilde{r}}=\frac{G}{2}\left[\frac{(1-G^2)}{\tilde{r}}+\tilde{r}\left(G^2\bigl(2\tilde{V}+\frac{\tilde{\omega}^2}{N^2}\tilde{\phi}^2\bigr)+\tilde{\phi}_{,\tilde{r}}^2+\tilde{\chi}_{,\tilde{r}}^2\right)\right],&
    \\
    \label{eq_eom_n2}
     &N_{,\tilde{r}}=\frac{N}{2}\left[\frac{(G^2-1)}{\tilde{r}}+\tilde{r}\left(G^2\left(\frac{\tilde{\omega}^2}{N^2}\tilde{\phi}^2-2\tilde{V}\right)+\tilde{\phi}_{,\tilde{r}}^2+\tilde{\chi}_{,\tilde{r}}^2\right)\right],&
     \\
     \label{eq_eom_phi2}
  & N^2 \tilde{\phi}_{,\tilde{r}\tilde{r}} + N^2 \left( 1 + G^2 - 2 \tilde{r}^2 G^2  \tilde{V} \right) \frac{\tilde{\phi}_{,\tilde{r}}}{\tilde{r}} + G^2\left( \tilde{\omega}^2 \tilde{\phi} - N^2 \frac{d\tilde{V}}{d\tilde{\phi}} \right) =0,&
   \\
   \label{eq_eom_chi2}
  & N^2 \tilde{\chi}_{,\tilde{r}\tilde{r}} + N^2 \left( 1 + G^2 -  2\tilde{r}^2 G^2  \tilde{V} \right) \frac{\tilde{\chi}_{,\tilde{r}}}{\tilde{r}}  -N^2G^2\frac{d\tilde{V}}{d\tilde{\chi}}=0.&
\end{align}
Then, these equations do not depend on $g$, one can treat the cases with different values of $g$ simultaneously.
This allows us to describe arbitrary values of $m_\phi$, even when considering $\chi_\mathrm{out} =\mathcal{O}(m_\mathrm{pl})$ in the following sections.
We solve Eqs.\eqref{eq_eom_g2}-\eqref{eq_eom_chi2} numerically under the boundary conditions \eqref{eq_bc_ori} and \eqref{eq_bc_inf} to obtain sequences of the stationary configurations for each $\chi_\mathrm{out}$ by using a shooting method.
\par
We introduce several physically meaningful quantities of the boson star.
Using the mass function
\begin{equation}
\label{eq_mass_def}
    M(r)=\frac{r}{2}\left(1-\frac{1}{G^2(r)}\right)m_\mathrm{pl}^2,
\end{equation}
the total mass is given by $M = M(\infty)$. 
We define compactness $C$ of the boson star as the maximum value of the function
\begin{equation}
\label{eq_compactness_def}
C(r) = \frac{M(r)}{m_\mathrm{pl}^2r},
\end{equation}
and define the radius $R$ of the boson star by the value of $r$ where 
$C(r)$ reaches its maximum. 
The justification for the definitions of radius $R$ and compactness $C$ is provided in the Appendix~\ref{sec_Definitions of the Radius and Compactness of Boson Stars}. 
Similarly to Eqs. \eqref{eq_scaling}, we introduce non-dimensional counterparts of $M$, $J$ and $R$ as
\begin{equation}
\label{eq_scaling_mn}
M = \frac{\sqrt{4\pi}m_\mathrm{pl}\tilde{M}}{g} , \quad J = \frac{4\pi\tilde{J}}{g^2},\quad R = \frac{\sqrt{4\pi}\tilde{R}}{gm_\mathrm{pl}}.
\end{equation}

\section{Initial State and Final State of Adiabatic Evolution}
\label{sec_Initial State and Final State of Adiabatic Evolution}
In this paper, we consider the scenario in which non-relativistic boson clouds evolve into compact boson stars through adiabatic change of an external parameter $\chi_{\rm out}$. 
To analyze the evolution, we specify the initial and final states.
Throughout the paper, quantities in the initial and final states are associated with subscripts $\mathrm{i}$ and $\mathrm{f}$, respectively.
\par
We first consider the initial state of adiabatic evolution. 
Non-relativistic boson clouds can spontaneously form through gravitational scattering.
As mentioned in Sec.~\ref{sec_Difficulty in the Formation of Compact Boson Stars Via Gravitational Scattering}, Ref.~\cite{Chan:2022bkz} calculated the evaporation rate $\Gamma$ for $\nu\gg 1$ and analyzed the threshold mass of whether a non-relativistic boson cloud either grows or evaporates.
We adopt the configuration at this threshold, as the initial state for the adiabatic evolution.
\par
The evaporation rate is given by
\begin{equation}
\label{eq_evap_rate}
    \Gamma = B \frac{(4\pi)^2 m_\phi^3 \rho_g^2}{m_\mathrm{pl}^4 k_g^6} \nu^2,
\end{equation}
as derived in Ref.~\cite{Chan:2022bkz,Chan:2023crj}, where $B\simeq3.5$.
The threshold is determined by the condition that the evaporation time (Eq.~\eqref{eq_evap_rate}) equals the relaxation time (Eq.~\eqref{eq_relaxationtime}). 
The value of $\nu$ at the threshold is
\begin{equation}
\label{starting_condition}
\nu_{\rm eq} =\sqrt{\frac{3\pi\ln(k_gL)}{4\sqrt{2}bB}}\simeq 1.5,
\end{equation}
where they adopted $\ln(k_gL)\simeq5$ and $b\simeq0.9$.
For $\nu > \nu_{\rm eq}$, the boson cloud evaporates, whereas for $\nu < \nu_{\rm eq}$, it grows, once it is spontaneously formed.
From Eq. \eqref{eq_ks_def}, the mass of the boson star at $\nu = \nu_{\rm eq}$ is
\begin{equation}
\label{eq_condi_initial}
M_{\rm eq} = \frac{\mu_0v}{4\pi\nu_{\rm eq}} \frac{m_\mathrm{pl}^2}{m_\phi}.
\end{equation}
The total particle number $J$ of the scalar field constituting the non-relativistic boson cloud is related to its mass $M$ by
\begin{equation}
J = \frac{M}{m_\phi}.
\end{equation}
Thus, the total particle number at $\nu = \nu_{\rm eq}$ is
\begin{equation}
\label{eq_condi_initial2}
J_{\rm eq} = \frac{\mu_0v}{4\pi\nu_{\rm eq}} \frac{m_\mathrm{pl}^2}{m_\phi^2}.
\end{equation}
A boson cloud with mass approximately $M \simeq M_{\rm eq}$ spontaneously forms on the timescale given by Eq.~\eqref{eq_relaxationtime}. 
However, if the boson cloud grows to a mass $M \gg M_{\rm eq}$, the evolution timescale becomes significantly longer than the formation timescale, making its growth within the age of the universe unlikely, as discussed in Sec.~\ref{sec_Difficulty in the Formation of Compact Boson Stars Via Gravitational Scattering}.
Therefore, the boson cloud with $\nu = \nu_{\rm eq}$, which satisfies $M=M_\mathrm{eq}$ and $J=J_\mathrm{eq}$, is adopted as the initial state for the adiabatic evolution.
Since the spatial gradient of the $\chi$-field can be assumed to be negligible for a non-relativistic boson cloud, the effective mass of the $\phi$-field is approximated by
\begin{equation}
m_\phi= g \chi_\mathrm{out}.
\end{equation}
Hence, the normalized mass $\tilde{M}_\mathrm{i}$ and particle number $\tilde{J}_\mathrm{i}$ at the initial state are expressed as
\begin{equation}
  \label{eq_start_n}
    \tilde{M}_\mathrm{i}
    =\frac{\mu_0 v}{4\pi\nu_{\rm eq}}\frac{1}{\tilde{\chi}_\mathrm{out,i}}
,\quad  \tilde{J}_\mathrm{i}
    =\frac{\mu_0 v}{4\pi\nu_{\rm eq}}\frac{1}{\tilde{\chi}_\mathrm{out,i}^2},
\end{equation}
where $\mu_0 v / (4\pi\nu_{\rm eq}) \simeq 1.4 \times 10^{-3}$.

\par
Next, we consider the final state of adiabatic evolution. 
We terminate the computation when the configuration becomes gravitationally unstable, where the total mass $M$ of the boson star reaches its maximum along the sequence of stationary solutions for a fixed $\tilde{\chi}_\mathrm{out}$. 
We show the total mass $M$ of stationary solutions for mini-boson stars with the Lagrangian given by Eq.~\eqref{eq_mini} in Fig.~\ref{fig_mo1plot}. 
The horizontal axis represents $1-\omega/m_\phi$, where $m_\phi$ is the mass of the $\phi$-field, and $\omega$ is the eigenvalue defined by Eq.~\eqref{eq_scalar}.
In mini-boson star cases, the stationary configurations form a one-parameter family. 
The blue point at $\omega/m_\phi=0.85$ marks the maximum mass of mini-boson stars and the critical point for gravitational stability.
The solid curve represents the stable branch, while the dashed curve represents the unstable branch \cite{PhysRevD.42.384, Hawley:2000dt, Gleiser:1988ih}. 
At the critical point, the compactness of the configuration $C$ is $\mathcal{O}(0.1)$.
\par
It is well-known that for fluid stars, such as neutron stars, the configuration with the maximum mass occurs at a specific central density, which coincides with the critical point for gravitational instability \cite{Shapiro:1983du}. 
This phenomenon is also observed in various boson star models and reflects a general feature of gravitational systems \cite{PhysRevLett.57.2485, Tamaki:2011zza, Kleihaus:2011sx}.
The configuration at the critical point is the most compact in the stable branch.

\section{Results}
\label{sec_results}
\subsection{A Series of the Stationary Configurations for each \texorpdfstring{$\chi_\mathrm{out}$}{chiout}}
\label{sec_Properties of the sequences of the stationary configurations for each chiout}
\begin{figure}[t]
  \centering
    \begin{subfigure}[b]{0.48\textwidth}
        \centering
        \includegraphics[width=\textwidth]{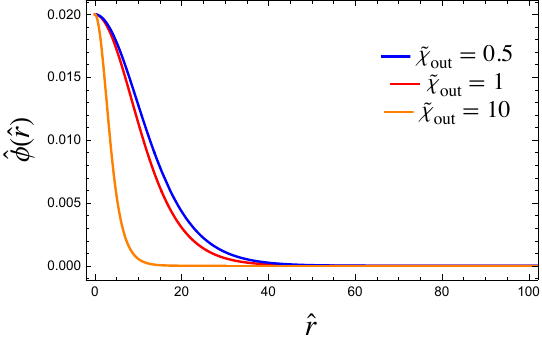}
        \protect\phantomcaption
        \label{fig_chioutpro_phi}
        \raisebox{0.5\height}{\hspace{1.3cm} \large \textbf{(a)}}
    \end{subfigure}
    \hfill
    \begin{subfigure}[b]{0.48\textwidth}
        \centering
        \includegraphics[width=\textwidth]{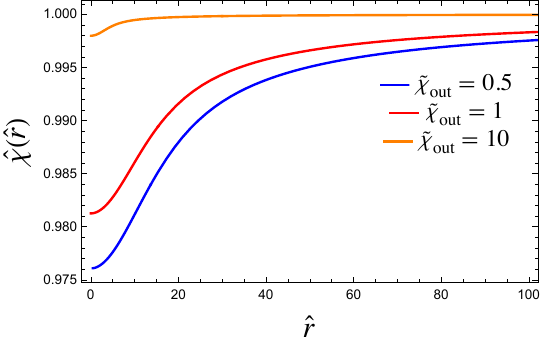}
         \protect\phantomcaption
        \label{fig_chioutpro_chi}
        \raisebox{0.5\height}{\hspace{1.3cm} \large \textbf{(b)}}
    \end{subfigure}
     \\
    \begin{subfigure}[b]{0.48\textwidth}
        \centering
        \includegraphics[width=\textwidth]{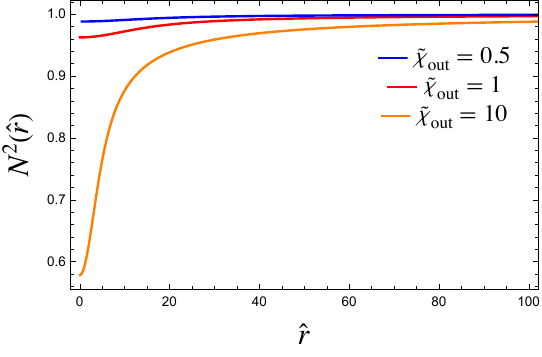}
        \protect\phantomcaption
        \label{fig_chioutpro_n}
        \raisebox{0.5\height}{\hspace{1.3cm} \large \textbf{(c)}}
    \end{subfigure}
    \hfill
    \begin{subfigure}[b]{0.48\textwidth}
        \centering
        \includegraphics[width=\textwidth]{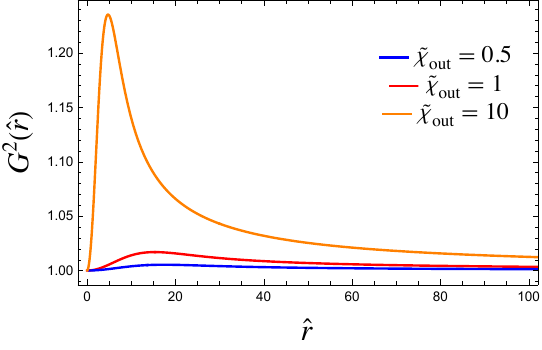}
        \protect\phantomcaption
        \label{fig_chioutpro_g}
        \raisebox{0.5\height}{\hspace{1.3cm} \large \textbf{(d)}}
    \end{subfigure}
  \caption{
Profiles of $\hat{\phi}(\hat{r})$, $\hat{\chi}(\hat{r})$, $N^2(\hat{r})$, and $G^2(\hat{r})$ of the stationary solutions for $\hat{\phi}_0=0.02$.
Here, we define $\hat{\phi}$, $\hat{\chi}$, $\hat{r}$ and $\hat{\omega}$ as given in Eqs.~\eqref{eq_hat_def}.
The blue, red, and orange curves represent the profiles for $\tilde{\chi}_\mathrm{out} = 0.5$, $1$, and $10$, respectively. 
When $\hat{\phi}_0$ is fixed to $0.02$, the eigenvalue $\omega$ becomes $1-\hat{\omega}=1.4\times10^{-2},1.8\times10^{-2}$ and $1.2\times10^{-1}$ for $\tilde{\chi}_\mathrm{out} = 0.5$, $1$, and $10$, respectively.
 }
  \label{fig_chioutpro}
\end{figure}

\begin{figure}[htbp]
\centering
  \begin{subfigure}[b]{1.0\textwidth}
        \centering
        \includegraphics[width=\textwidth]{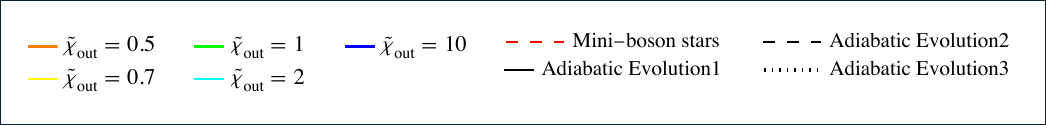}
    \end{subfigure}
     \\
       \vspace{1em} 
    \begin{subfigure}[b]{0.48\textwidth}
        \centering
        \begin{overpic}[width=\textwidth]{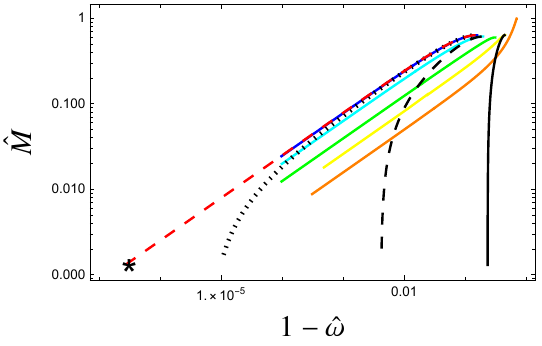}
         \put(20,55){\textbf{(a)}} 
        \end{overpic}
        \protect\phantomcaption
        \label{fig_mphi0}
    \end{subfigure}
    \hfill
    \begin{subfigure}[b]{0.48\textwidth}
        \centering
        \begin{overpic}[width=\textwidth]{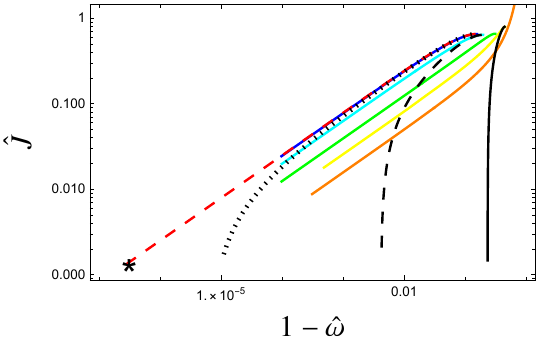}
         \put(20,55){\textbf{(b)}} 
        \end{overpic}
        \protect\phantomcaption
        \label{fig_nphi0}
    \end{subfigure}
     \\
    \begin{subfigure}[b]{0.48\textwidth}
        \centering
        \begin{overpic}[width=\textwidth]{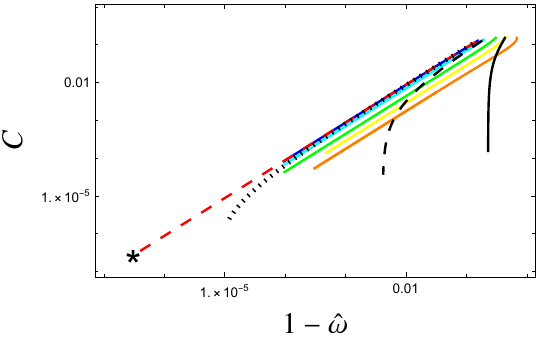}
         \put(20,55){\textbf{(c)}} 
        \end{overpic}
         \protect\phantomcaption
        \label{fig_cphi0}
    \end{subfigure}
    \hfill
   \begin{subfigure}[b]{0.48\textwidth}
        \centering
        \begin{overpic}[width=\textwidth]{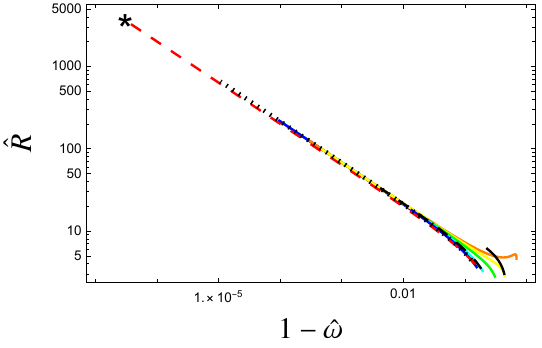}
         \put(85,55){\textbf{(d)}} 
        \end{overpic}
        \protect\phantomcaption
        \label{fig_rphi0}
    \end{subfigure}
    \hfill
   \begin{subfigure}[b]{0.48\textwidth}
        \centering
        \begin{overpic}[width=\textwidth]{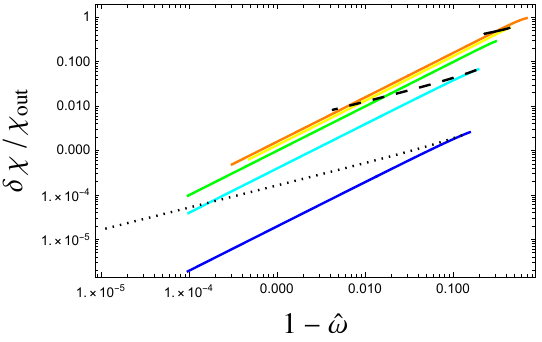}
         \put(20,55){\textbf{(e)}} 
        \end{overpic}
        \protect\phantomcaption
        \label{fig_dchi0phi0}
    \end{subfigure}
    \hfill
    \begin{subfigure}[b]{0.48
 \textwidth}
        \centering
        \begin{overpic}[width=\textwidth]{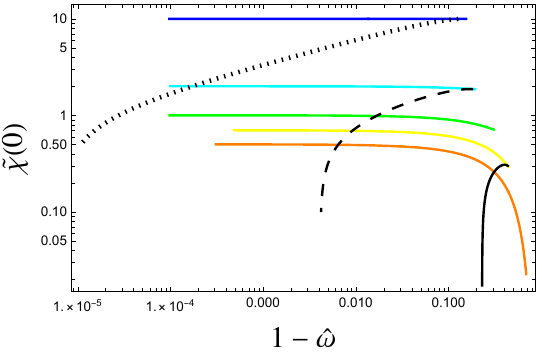}
         \put(85,55){\textbf{(f)}} 
        \end{overpic}
        \protect\phantomcaption
        \label{fig_chi0phi0}
    \end{subfigure}
  \caption{
The normalized total mass $\hat{M}$, total particle number $\hat{J}$, compactness $C$, radius $\hat{R}$, central $\chi$-field value $\tilde{\chi}(0)$, and magnitude of the gradient of $\chi$-field $\delta\chi/\chi_\mathrm{out}$ of stationary solutions as functions of the eigenvalue $\hat{\omega}$ for various values of $\tilde{\chi}_\mathrm{out}$, where $\delta\chi=\chi_\mathrm{out}-\chi(0)$.
Here, we define $\hat{M}$, $\hat{J}$, $C$ and $\hat{R}$ as given in Eqs.~\eqref{eq_hat_def2}.
These are plotted up to the final states where $\hat{M}$ and $\hat{J}$ reach their maximum values for each $\chi_\mathrm{out}$.  
The horizontal axis represents $1-\hat{\omega}$, with smaller values corresponding to non-relativistic configurations and values near 1 indicating relativistic ones.
In each panel, the dotted red curves represent the quantities for mini-boson stars. 
These quantities are defined using Eq.~\eqref{eq_mani_dimensionless1} and Eq.~\eqref{eq_mani_dimensionless2}.
The solid, dashed, and dotted curves, labeled Adiabatic Evolution 1, 2, and 3, represent the adiabatic evolution paths from the initial states defined by Eq.~\eqref{eq_start_n} to the final states corresponding to $\tilde{\chi}_\mathrm{out,f} \simeq 0.7$, 2, and 10, respectively.
The black asterisks represent the initial state of mini-boson stars.
Note that non-relativistic regime of mini-boson stars can be described by Shr\"{o}dinger-Poisson equations\cite{PhysRevD.42.384}.
}
  \label{fig_nmphi0}
\end{figure}

Solving Eqs.~\eqref{eq_eom_g2}-\eqref{eq_eom_chi2} numerically under the boundary conditions~\eqref{eq_bc_ori} and~\eqref{eq_bc_inf}, we obtain sequences of the stationary configurations for each $\chi_\mathrm{out}$. 
First, we examine the dependence on $\chi_\mathrm{out}$.
The effective mass of the $\phi$-field, given by Eq.~\eqref{mphi}, depends on the value of $\chi_\mathrm{out}$.
In this section, we discuss the results using the following dimensionless variables:
\begin{equation}
\label{eq_hat_def}
\hat{\omega}=\tilde{\omega}/\tilde{\chi}_\mathrm{out},\quad\hat{r}=\tilde{\chi}_\mathrm{out}\tilde{r},\quad\hat{\phi}^2=\tilde{\phi}^2/\tilde{\chi}_\mathrm{out}^2,\quad\hat{\chi}^2=\tilde{\chi}^2/\tilde{\chi}_\mathrm{out}^2,
\end{equation}
\begin{equation}
\label{eq_hat_def2}
\hat{M}=\tilde{M}\tilde{\chi}_\mathrm{out},\quad\hat{J}=\tilde{J}\tilde{\chi}^2_\mathrm{out},\quad\hat{R}=\tilde{R}\tilde{\chi}_\mathrm{out},
\end{equation}
normalized by $\tilde{\chi}_\mathrm{out}$, which corresponds to normalization by $m_\phi(\infty)$.
In the case of mini-boson stars, we also introduce the following dimensionless variables normalized by $m_\phi$:
\begin{equation}
\label{eq_mani_dimensionless1}
\hat{\omega} = \frac{\omega}{m_\phi}, \quad \hat{r} = r m_\phi, 
\end{equation}
as well as
\begin{equation}
\label{eq_mani_dimensionless2}
\hat{M} = \frac{M m_\phi}{m_\mathrm{pl}^2}, \quad 
\hat{J} = \frac{J m_\phi^2}{m_\mathrm{pl}^2}, \quad 
\hat{R} = R m_\phi.
\end{equation}
\par
We show profiles of $\hat{\phi}(\hat{r})$, $\hat{\chi}(\hat{r})$, $N^2(\hat{r})$, and $G^2(\hat{r})$ of the stationary solutions for $\hat{\phi}_0=0.02$ in Fig.~\ref{fig_chioutpro}. 
In each panel, the blue, red, and orange curves represent the profiles for $\tilde{\chi}_\mathrm{out} = 0.5$, $1$, and $10$, respectively. 
When $\hat{\phi}_0$ is fixed to $0.02$, the eigenvalue $\omega$ becomes $1-\hat{\omega}=1.4\times10^{-2},1.8\times10^{-2}$ and $1.2\times10^{-1}$ for $\tilde{\chi}_\mathrm{out} = 0.5$, $1$, and $10$, respectively.
The quantity $1-\hat{\omega}$ represents the binding energy of the $\phi$-field per unit mass. 
From Fig.~\ref{fig_chioutpro}, we observe that the behavior of stationary solutions changes significantly depending on $\chi_\mathrm{out}$. 
For larger $\chi_\mathrm{out}$, the profiles of $N^2(\hat{r})$ and $G^2(\hat{r})$ become steeper, reflecting stronger self-gravity of the boson star (Fig.~\ref{fig_chioutpro_n}, Fig.~\ref{fig_chioutpro_g}).  
In contrast, for smaller $\chi_\mathrm{out}$, the spatial gradient of $\hat{\chi}(\hat{r})$ becomes steeper (Fig.~\ref{fig_chioutpro_chi}).
In our model, the binding energy is composed of contributions from self-gravity and the reduction of the mass of the $\phi$-field due to the spatial gradient of the $\chi$-field.  
These results demonstrate that, for a fixed $\hat{\phi}_0$, smaller $\chi_\mathrm{out}$ values results in a larger contribution of the $\chi$-field's gradient to the total binding energy of the $\phi$-field.

\par
We show the normalized quantities: the total mass $\hat{M}$, total particle number $\hat{J}$, compactness $C$, radius $\hat{R}$, central $\chi$-field value $\tilde{\chi}(0)$, and magnitude of the gradient of $\chi$-field $\delta\chi/\chi_\mathrm{out}$ of stationary solutions as functions of the eigenvalue $\hat{\omega}$ for various values of $\tilde{\chi}_\mathrm{out}$ in Fig.~\ref{fig_nmphi0}, where $\delta\chi=\chi_\mathrm{out}-\chi(0)$.
These are plotted up to the final states where $\hat{M}$ and $\hat{J}$ reach their maximum values for each $\chi_\mathrm{out}$. 
The horizontal axis represents $1-\hat{\omega}$, with smaller values corresponding to non-relativistic configurations and values near 1 indicating relativistic ones.
In each panel, the red curves show the quantities for mini-boson stars.
These quantities are defined using Eq.~\eqref{eq_mani_dimensionless1} and Eq.~\eqref{eq_mani_dimensionless2}.
The solid, dashed, and dotted curves, labeled Adiabatic Evolution 1, 2, and 3, represent the adiabatic evolution paths from the initial states defined by Eq.~\eqref{eq_start_n} to the final states corresponding to $\tilde{\chi}_\mathrm{out,f} \simeq 0.7$, 2, and 10, respectively.
The black asterisks represent the initial state of mini-boson stars.
Note that non-relativistic regime of mini-boson stars can be described by Shr\"{o}dinger-Poisson equations\cite{PhysRevD.42.384}.
\par
From Fig.~\ref{fig_mphi0}-Fig.~\ref{fig_rphi0}, we observe that as $\tilde{\chi}_\mathrm{out}$ increases, the stationary solutions resemble mini-boson stars.
For $\tilde{\chi}_\mathrm{out} > \mathcal{O}(1)$, the sequences of solutions align closely with the dotted red curve representing mini-boson stars.
At the critical points, the values of $\hat{J}$ and $\hat{M}$ nearly match those of mini-boson stars.
In contrast, as $\tilde{\chi}_\mathrm{out}$ decreases, deviations from mini-boson stars become more pronounced. 
The mass $\hat{M}$, total particle number $\hat{J}$, and compactness $C$ are all smaller than those of mini-boson stars with the same $\hat{\omega}$. 
For $\tilde{\chi}_\mathrm{out} < \mathcal{O}(1)$, the critical points occur at smaller $\hat{\omega}$ values than those of mini-boson stars. 
At these critical points, $\hat{M}$, $\hat{J}$, and $\hat{R}$ exceed the corresponding mini-boson star values.
Regardless of the value of $\tilde{\chi}_\mathrm{out}$, the compactness at the critical point remains nearly constant at $C \simeq 0.15$.
\par
From Fig.~\ref{fig_cphi0}, we observe that decreasing $\omega$ while keeping $\chi_\mathrm{out}$ fixed increases the compactness $C$.
Combined with Fig.~\ref{fig_cphi0} - Fig.~\ref{fig_dchi0phi0}, this suggests that, for fixed $\chi_\mathrm{out}$, a more compact boson star corresponds to a larger value of $\delta\chi / \chi_\mathrm{out}$.
For $\tilde{\chi}_\mathrm{out} \gg \mathcal{O}(1)$, the gradient $\delta\chi / \chi_\mathrm{out}$ remains small across all $\hat{\omega}$ values, leading to $\chi(0) \simeq \chi_\mathrm{out}$.
In contrast, for $\tilde{\chi}_\mathrm{out} \lesssim \mathcal{O}(1)$ as $\hat{\omega}$ decreases, $\delta\chi/\chi_\mathrm{out}$ becomes significantly large, indicating a notable deviation from the $\chi(r) \simeq \chi_\mathrm{out}$ approximation.
Namely, even if $\chi_\mathrm{out}$ increases, $\chi(0)$ may remain small.
In this regime, the boson star resembles a non-topological soliton star, which represents a localized configuration bound even in the absence of gravity \cite{Endo:2022uhe, Tamaki:2011zza, Collodel_2022}.
The binding in this case arises due to the large gradient of the $\chi$-field, which traps the $\phi$-field and forms a bound state.
This trend is consistent with Fig.~\ref{fig_chioutpro_chi}, where a smaller $\tilde{\chi}_\mathrm{out}$ leads to a significant spatial gradient of the $\chi$-field.
In the present model, we consider the scenario where the $\phi$-field initially forms a boson cloud through gravitational scattering within the spatially homogeneous $\chi$-field. 
Therefore, we do not assume that the spatial gradient of the $\chi$-field is sufficiently large in the non-relativistic boson cloud regime. 
This assumption is consistent with the condition that the $\chi$-field remains in slow-roll motion, as will be discussed in Sec.~\ref{sec_Discussion and Conclusion}.
Note that the initial configurations of Adiabatic Evolution 1, 2, and 3 significantly differ from the initial states that well-approximated by the Schrödinger-Poisson system.
However, since the mechanism for the formation of seeds under strong interactions with the $\chi$-field remains unclear, we adopt this assumption in this paper.

\subsection{Adiabatic Evolution Paths and Mechanisms}
\label{sec_Adiabatic Evolution Paths and Mechanisms}

\begin{figure}[t]
  \centering
    \begin{subfigure}[b]{0.48\textwidth}
        \centering
        \includegraphics[width=\textwidth]{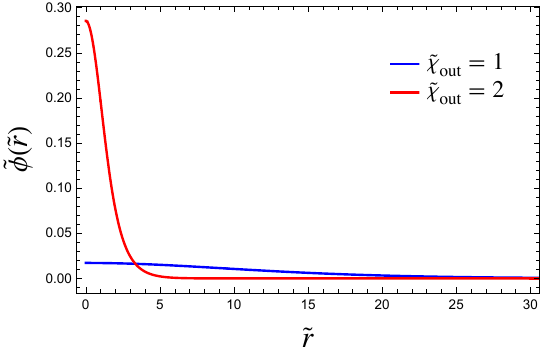}
        \protect\phantomcaption
        \label{fig_adipro_phi}
        \raisebox{0.5\height}{\hspace{1.1cm} \large \textbf{(a)}}
    \end{subfigure}
    \hfill
    \begin{subfigure}[b]{0.48\textwidth}
        \centering
        \includegraphics[width=\textwidth]{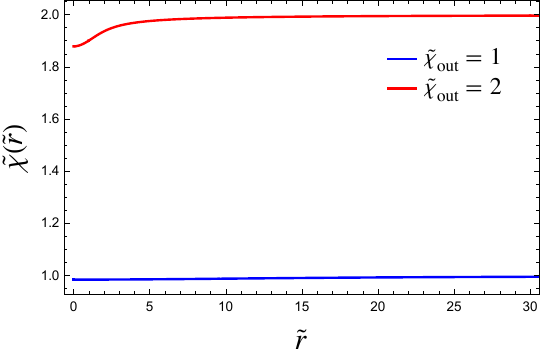}
        \protect\phantomcaption
        \label{fig_adipro_chi}
        \raisebox{0.5\height}{\hspace{1.1cm} \large \textbf{(b)}}
    \end{subfigure}
     \\
    \begin{subfigure}[b]{0.48\textwidth}
        \centering
        \includegraphics[width=\textwidth]{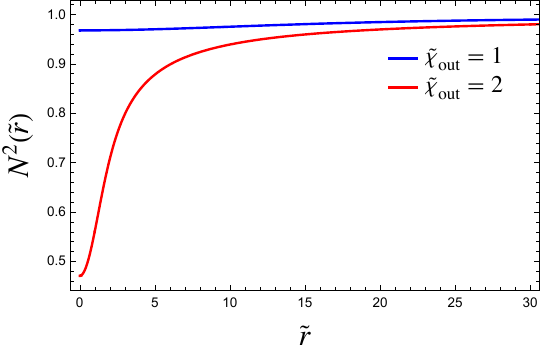}
        \protect\phantomcaption
        \label{fig_adipro_n}
        \raisebox{0.5\height}{\hspace{1.1cm} \large \textbf{(c)}}
    \end{subfigure}
    \hfill
    \begin{subfigure}[b]{0.48\textwidth}
        \centering
        \includegraphics[width=\textwidth]{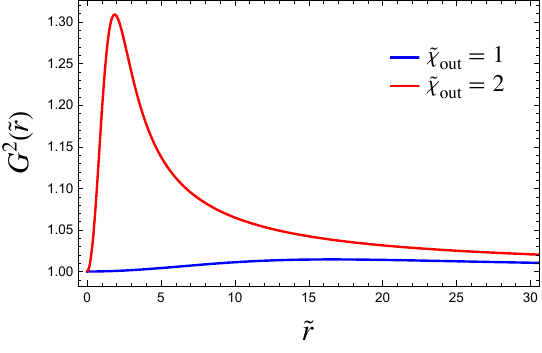}
        \protect\phantomcaption
        \label{fig_adipro_g}
        \raisebox{0.5\height}{\hspace{1.1cm} \large \textbf{(d)}}
    \end{subfigure}
  \caption{
  Profiles of $\tilde{\phi}(\tilde{r})$, $\tilde{\chi}(\tilde{r})$, $N^2(\tilde{r})$, and $G^2(\tilde{r})$ of the stationary solutions for a normalized total particle number $\tilde{J}=1.6\times10^{-1}$.
  These profiles are in the adiabatic evolution path 1 shown in Fig.~\ref{fig_nmphi0}. 
  The blue, and red curves represent the profiles for $\tilde{\chi}_\mathrm{out} = 1$, and 2, respectively. 
  Each horizontal axis is $\tilde{r}\equiv rgm_\mathrm{pl}/\sqrt{4\pi}$. Note that $m_\phi$, the effective mass of $\phi$-field, varies in the adiabatic evolution.
  }
  \label{fig_adipro}
\end{figure}

\begin{figure}[t]
\centering
\includegraphics[width=0.8\linewidth]{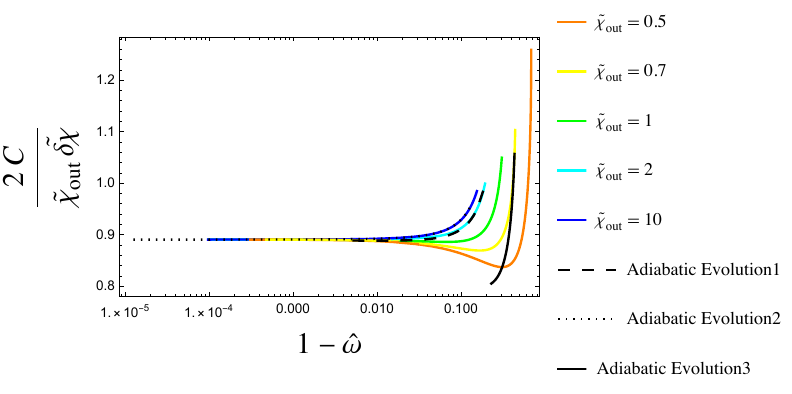}
\caption{
The ratio of the binding energy associated with the $\chi$-field gradient to the gradient energy, given by $2 C/\tilde{\chi}_{\mathrm{out}} \delta\tilde{\chi}$, as a function of $1 - \hat{\omega}$. 
Similarly to Fig.~\ref{fig_nmphi0}, each colored curve represents a sequence of stationary solutions for each $\chi_{\mathrm{out}}$. 
The solid, dashed, and dotted curves correspond to the three different adiabatic evolution paths.
}
\label{fig_bindingcheck}
\end{figure}

Next, we examine how the stationary solutions change along the adiabatic evolution paths, where $\chi_\mathrm{out}$ is varied while keeping the total particle number $J$ constant.
It is important to note that as $\chi_\mathrm{out}$ changes, $\hat{J}$, the particle number normalized by $g\chi_\mathrm{out}$, does not conserve.
\par
In Fig.~\ref{fig_nmphi0}, the solid, dashed and dotted curves, labeled Adiabatic Evolution 1, 2, and 3, represent adiabatic evolution paths with $\tilde{\chi}_\mathrm{out,f} \simeq 0.7$, 2, and 10, respectively. 
The black asterisks represent the initial state of mini-boson stars.
From these adiabatic evolution paths, we observe that $\chi_\mathrm{out}$ varies along the trajectory. 
As shown in Fig.~\ref{fig_cphi0}, boson stars become more compact, as $\chi_\mathrm{out}$ increases and the central value of the $\chi$-field, $\chi(0)$, also increases (Fig.~\ref{fig_chi0phi0}).
This implies that $m_\phi(0)$ becomes larger.
Furthermore, during the evolution, the total particle number $J$ is conserved, while the total mass $M$ increases and its radius $R$ decreases, as demonstrated in Fig.~\ref{fig_mphi0} and Fig.~\ref{fig_rphi0}.
\par
 We show the profiles of $\tilde{\phi}(\tilde{r})$, $\tilde{\chi}(\tilde{r})$, $N^2(\tilde{r})$, and $G^2(\tilde{r})$ of the stationary solutions for a normalized total particle number $\tilde{J}=1.6\times10^{-1}$ in Fig.~\ref{fig_adipro}.
These profiles are in Adiabatic Evolution 1 shown in Fig.~\ref{fig_nmphi0}. 
The blue, and red curves represent the profiles for $\tilde{\chi}_\mathrm{out} = 1$, and 2, respectively. 
The horizontal axis is $ \tilde{r}\equiv rgm_\mathrm{pl}/\sqrt{4\pi}$. Note that $m_\phi$, the effective mass of $\phi$-field, varies in the adiabatic evolution.
From Fig.~\ref{fig_adipro}, we observe that for a larger $\chi_\mathrm{out}$ while keeping $J$ fixed, the gravitational potential $N^2(\tilde{r})$ and $G^2(\tilde{r})$ become deeper, indicating the self-gravity of the boson star becomes stronger, as shown in Fig.~\ref{fig_adipro_n} and Fig.~\ref{fig_adipro_g}.
Moreover, as $\chi_\mathrm{out}$ increases, the inner value of the $\chi$-field also grows, as illustrated in Fig.~\ref{fig_adipro_chi}.
\par
Comparing the three adiabatic evolution paths shown in Fig.~\ref{fig_nmphi0}, we find that configurations deviate more significantly from the behavior of mini-boson stars as $\chi_\mathrm{out,f}$ decreases (or equivalently, as $J$ increases, or as $g$ increases when $m_{\phi,\mathrm{f}}$ is fixed).
This deviation arises because a larger spatial gradient of the $\chi$-field leads to a larger spatial gradient of $m_\phi$. 
As discussed earlier, it is shown that a smaller $\chi_\mathrm{out}$ leads to a larger spatial gradient of the $\chi$-field for a fixed $\omega$ (Fig.~\ref{fig_dchi0phi0}). 
Conversely, for a fixed $J$, a larger $\chi_{\mathrm{out}}$ results in an increased gravitational potential, which in turn enhances the spatial gradient of the $\chi$-field (Fig.~\ref{fig_adipro_chi}).
\par
To investigate the role of $\chi_{\mathrm{out}}$ and $C$ in determining the spatial gradient of the $\chi$-field, we consider a virial relation for the $\chi$-field, analogous to a harmonic oscillator system. 
The ratio of the binding energy associated with the $\chi$-field gradient ($\simeq M\delta m_\phi / m_\phi$) to the gradient energy ($\simeq (\delta \chi)^2 R/2$) is expected to satisfy
\begin{equation}
    \label{eq_chi_virial}
    \frac{M\delta m_\phi/m_\phi}{(\delta\chi)^2R/2} = \frac{2m_\mathrm{pl}^2 C}{\chi_{\mathrm{out}} \delta\chi} \simeq 1.
\end{equation}
Here, we use the relations $m_\phi = g \chi_{\mathrm{out}}$ and $\delta m_\phi = g \delta \chi$.
In Fig.~\ref{fig_bindingcheck}, we plot this ratio as a function of $1 - \hat{\omega}$. 
Similarly to Fig.~\ref{fig_nmphi0}, each colored curve represents a sequence of stationary solutions for each $\chi_{\mathrm{out}}$. 
The solid, dashed, and dotted curves correspond to the three different adiabatic evolution paths.
From Fig.~\ref{fig_bindingcheck}, we confirm that Eq.~\eqref{eq_chi_virial} holds. 
Therefore, this result implies that the gradient energy of the $\chi$-field is equal to the binding energy associated with the $\chi$-field gradient.
\par
Eq.~\eqref{eq_chi_virial} also implies that $\delta\chi/\chi_{\mathrm{out}}$ can be expressed as
\begin{equation}
    \label{eq_deltachi}
    \frac{\delta\chi}{\chi_{\mathrm{out}}} \simeq \frac{m_\mathrm{pl}^2 C}{\chi_{\mathrm{out}}^2}.
\end{equation}
This relation indicates that $\delta\chi/\chi_{\mathrm{out}}$ increases for larger $C$ or smaller $\chi_{\mathrm{out}}$.
From Fig.~\ref{fig_cphi0}, we observe that for a fixed $\chi_{\mathrm{out}}$, compactness follows $C \propto 1 - \hat{\omega}$. 
Additionally, Fig.~\ref{fig_nphi0} shows that the total particle number satisfies $J m_\phi^2 / m_\mathrm{pl}^2 \propto (1 - \hat{\omega})^{1/2}$. 
Combining these results, the scaling of the $\chi$-field gradient is given by
\begin{equation}
    \frac{\delta\chi}{\chi_{\mathrm{out}}} \propto
    \begin{cases}
        (1 - \hat{\omega})^{1/2}, & \text{for fixed } J, \\
        1 - \hat{\omega}, & \text{for fixed } \chi_{\mathrm{out}}.
    \end{cases}
\end{equation}
This scaling is consistent with the adiabatic evolution paths and constant-$\chi_{\mathrm{out}}$ curves in Fig.~\ref{fig_dchi0phi0}; 
The gradient of the $\chi$-field during adiabatic evolution changes more slowly compared to the case with a constant $\chi_\mathrm{out}$.
Thus, we can conclude that the gradient does not change much throughout the evolution, and the acquisition of the $\chi$-field gradient is primarily determined by the choice of initial conditions rather than by the increase in compactness during adiabatic evolution.
Furthermore, the ratio of the binding energy due to the $\chi$-field gradient $M\delta m_\phi /m_\phi$ to the gravitational one $M^2/(m_\mathrm{pl}^2R)$ is given by
\begin{equation}
    \label{eq_binding_rate}
    \frac{\delta m_\phi / m_\phi}{C} = \frac{\delta\chi}{\chi_{\mathrm{out}} C} \simeq \frac{m_\mathrm{pl}^2}{\chi_{\mathrm{out}}^2},
\end{equation}
where Eq.~\eqref{eq_deltachi} has been used in the second equality.
According to Eq.~\eqref{eq_binding_rate}, when this ratio is significantly greater than unity, the interaction with the $\chi$-field dominates the self-gravity, which happens when $\chi_{\mathrm{out}}$ is much smaller than the Planck scale, particularly at the initial stage.

\par
\begin{figure}[t]
  \centering
    \begin{subfigure}[b]{0.47\textwidth}
        \centering
        \includegraphics[width=\textwidth]{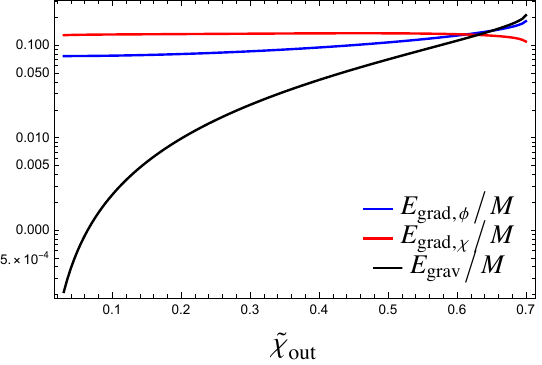}
        \caption{Adiabatic Evolution 1 ($\tilde{\chi}_\mathrm{out,f}\simeq0.7$)}
        \label{fig_ploerate1}
    \end{subfigure}
    \hfill
    \hfill
    \begin{subfigure}[b]{0.48\textwidth}
        \centering
        \includegraphics[width=\textwidth]{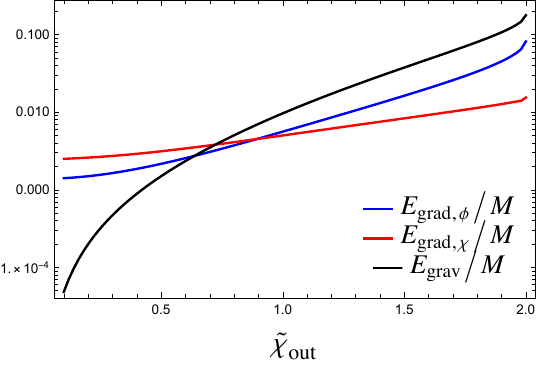}
        \caption{Adiabatic Evolution 2 ($\tilde{\chi}_\mathrm{out,f}\simeq2$)}
        \label{fig_ploerate2}
    \end{subfigure}
     \hfill
     \vspace{1em} 
    \begin{subfigure}[b]{0.48\textwidth}
        \centering
        \includegraphics[width=\textwidth]{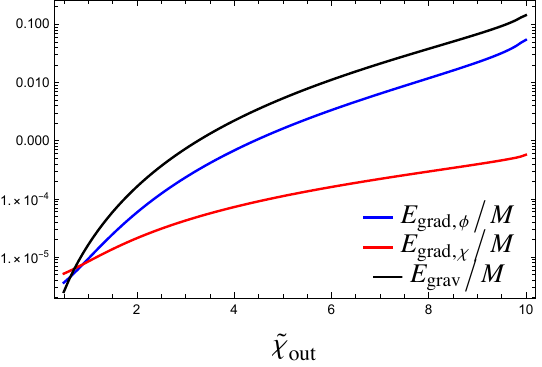}
        \caption{Adiabatic Evolution 3 ($\tilde{\chi}_\mathrm{out,f}\simeq10$)}
        \label{fig_ploerate3}
    \end{subfigure}
  \caption{
The ratios of the quantities, defined by Eq.~\eqref{eq_mass_grad_def}-\eqref{eq_mass_newton_def}, to $M$ along adiabatic the evolution path.
 Panels (a), (b), and (c) correspond to Adiabatic Evolution 1 ($\tilde{\chi}_\mathrm{out,f}\simeq0.7$), Adiabatic Evolution 2 ($\tilde{\chi}_\mathrm{out,f}\simeq1$), Adiabatic Evolution 3 ($\tilde{\chi}_\mathrm{out,f}\simeq10$), respectively, shown in Fig.~\ref{fig_nmphi0}. 
 The blue, red, and black curves represent $E_\mathrm{grad,\phi}/M$, $E_\mathrm{grad,\chi}/M$, and $E_\mathrm{grav}/M$, respectively. 
  }
  \label{fig_ploerate}
\end{figure}

Using Eq.~\eqref{eq_eom_g}, the total mass, defined in Eq.~\eqref{eq_mass_def}, can be expressed as  
\begin{equation}
\label{eq_mass_tot}
   M= -4\pi\int r^2{T^0}_0dr.
\end{equation}  
The contributions of the gradient energies of the $\phi$-field and $\chi$-field to $M$ are defined as
\begin{equation}
\label{eq_mass_grad_def}
    E_{\mathrm{grad},\phi}=4\pi\int r^2 \frac{1}{2G^2(r)}\phi'^2(r)dr ,\quad
    E_\mathrm{grad,\chi}=4\pi\int r^2 \frac{1}{2G^2(r)}\chi'^2(r)dr.
\end{equation}
The contribution of gravitational binding can be estimated by 
\begin{equation}
\label{eq_mass_newton_def}
    E_\mathrm{grav}=\left|\frac{M^2}{Rm_\mathrm{pl}^2}\right|,
\end{equation}
where $R$ is the radius of the boson star, as defined in the previous section.
We show the ratios of the quantities, defined by Eq.~\eqref{eq_mass_grad_def}~-~\eqref{eq_mass_newton_def}, to $M$ along the adiabatic evolution paths in Fig.~\ref{fig_ploerate}.
 Panels (a), (b), and (c) correspond to Adiabatic Evolution 1 ($\tilde{\chi}_\mathrm{out,f}\simeq0.7$), Adiabatic Evolution 2 ($\tilde{\chi}_\mathrm{out,f}\simeq1$), Adiabatic Evolution 3 ($\tilde{\chi}_\mathrm{out,f}\simeq10$), respectively, shown in Fig.~\ref{fig_nmphi0}. 
 The blue, red, and black curves represent $E_\mathrm{grad,\phi}/M$, $E_\mathrm{grad,\chi}/M$, and $E_\mathrm{grav}/M$, respectively. 
From Fig.~\ref{fig_ploerate3}, we observe that in Adiabatic Evolution 3, $E_\mathrm{grad,\chi}$ is significantly smaller than $E_\mathrm{grav}$, except at the initial stage.  
In contrast, as \(\tilde{\chi}_\mathrm{out,f}\) decreases, $E_\mathrm{grad,\chi}$ becomes much larger than $E_\mathrm{grav}$.
Notably, in  Adiabatic Evolution 1, $E_\mathrm{grad,\chi}$ dominates in most regions, as shown in Fig.~\ref{fig_ploerate1}.  
From these results, we find that configurations during adiabatic evolution behave consistently with Eq.~\eqref{eq_binding_rate}; these configurations with $\chi_\mathrm{out}\ll m_\mathrm{pl}$ resemble non-topological solitons, in which the gradient energy of the $\chi$-field balances with the binding energy through the interaction potential.
Conversely, in a gravitationally bound system, the Virial relation $2E_\mathrm{grad,\phi} \simeq E_\mathrm{grav}$ holds. 
Furthermore, the contribution of the $\chi$-field gradient energy to the total mass remains nearly constant throughout the adiabatic evolution, while the contribution from gravitational energy increases as $\chi_\mathrm{out}$ increases, as expected.
\par
In this section, we confirmed that boson stars can become more compact through the slow-roll motion of the $\chi$-field, irrespectively of the final effective mass of the $\phi$-field. 
Compactness increases as the $\chi$-field evolves to larger values, corresponding to an increase of the effective mass of the $\phi$-field. 
During this process, the energy of the slow-roll motion of the $\chi$-field is transferred to the rest mass energy of the $\phi$-field.
Inside the star, the increase in the rest mass enhances gravitational binding energy, which further amplifies the gradient energy of the $\phi$-field and $\chi$-field. 
As a result, the total mass $M$ increases while the radius $R$ decreases, leading to an increase in the compactness $C$.
On the other hand, when the $\chi$-field evolves in the opposite direction, in which the effective mass $m_\phi$ decreases (i.e., the $\chi$-field value becomes smaller in this model), the compactness decreases.
In this case, the reduction of $m_\phi$ weakens the self-gravity of the system.

\subsection{Evaluation of the Change in Adiabatic Evolution}
\label{sec_Evaluation of the Change in Adiabatic Evolution}

In the previous section, we showed that the boson star becomes more compact through adiabatic evolution as $\chi_\mathrm{out}$ increases. 
Here, we investigate the magnitude of the change in $\chi_\mathrm{out}$, defined as  
\begin{equation}
\label{eq_delta_chiout}
    \Delta \chi_\mathrm{out}= \chi_\mathrm{out,f}-\chi_\mathrm{out,i},
\end{equation}   
required for the non-relativistic boson cloud to become sufficiently compact.
As previously discussed, we take the threshold for the evaporation of a boson cloud as the initial state and evolve it until the gravitational instability sets in.
For solution paths with $\tilde{\chi}_\mathrm{out,f} \ll \mathcal{O}(1)$, the binding energy due to the gradient of the $\chi$-field dominates at the initial stage.
However, in the present scenario, the initial state is to be prepared as a solution derived from the Schrödinger-Poisson system, where the gradient of the $\chi$-field is negligible.
Thus, we restrict our consideration to the adiabatic evolution in the region where $\tilde{\chi}_\mathrm{out,f} \gtrsim \mathcal{O}(1)$.
The validity of this assumption will be discussed in Sec.~\ref{sec_Discussion and Conclusion}.
The total particle number $J$ remains constant throughout the adiabatic evolution.

\begin{figure}[t]
\centering
\includegraphics[width=0.6\linewidth]{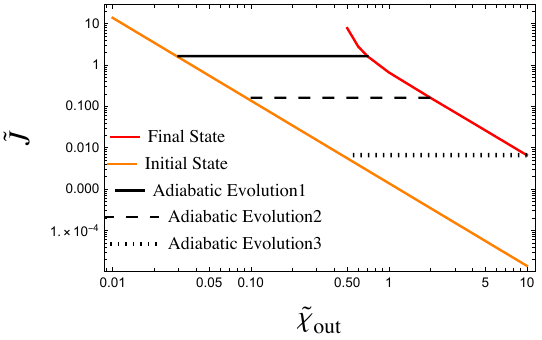}
\caption{
Adiabatic evolution paths in the $\tilde{\chi}_\mathrm{out}$–$\tilde{J}$ plane.
The orange line represents the initial state defined by Eq.~\eqref{eq_start_n}.
The red curve represents the final states corresponding to the maximum total particle numbers for each $\chi_\mathrm{out}$. 
The solid, dashed, and dotted lines represent the same adiabatic evolution paths shown in Fig.~\ref{fig_nmphi0}, terminating at $\tilde{\chi}_\mathrm{out,f} \simeq 0.7,2$, and 10, respectively.  
}
\label{fig_n-g_graph}
\end{figure}

\par
We show adiabatic evolution paths in the $\tilde{\chi}_\mathrm{out}$–$\tilde{J}$ plane in Fig.~\ref{fig_n-g_graph}.
The orange line represents the initial state defined by Eq.~\eqref{eq_start_n}.
The red curve represents the final states corresponding to the maximum total particle numbers for each $\chi_\mathrm{out}$. 
We observe that, for $\tilde{\chi}_\mathrm{out,f} \gtrsim \mathcal{O}(1)$, the total particle number in the final state, $\tilde{J}_\mathrm{f}$, follows the same scaling relation as in the initial state, $\tilde{J} \propto \tilde{\chi}^{-2}_\mathrm{out}$. 
In contrast, for $\tilde{\chi}_\mathrm{out,f} \lesssim \mathcal{O}(1)$, the dependence of $\tilde{J}_\mathrm{f}$ on $\tilde{\chi}_\mathrm{out,f}$ becomes stronger.
These trends are also shown in Fig.~\ref{fig_nphi0}.
The solid, dashed and dotted lines in Fig.~\ref{fig_n-g_graph} represent the same adiabatic evolution paths as shown in Fig.~\ref{fig_nmphi0}, with $\tilde{\chi}_\mathrm{out,f} \simeq 0.7,2$, and 10, respectively.

\par  
We evaluate $\Delta \chi_\mathrm{out}$, which represents the required change in $\chi_\mathrm{out}$ for a non-relativistic boson cloud to evolve into a compact boson star.
From Fig.~\ref{fig_nphi0}, we find that $\hat{J}_\mathrm{f} = \mathcal{O}(10^{-1})$ in the region where $\tilde{\chi}_\mathrm{out,f} \gtrsim \mathcal{O}(1)$. 
Therefore, the normalized total particle number $\tilde{J}_\mathrm{f}$ can be expressed as 
\begin{equation}
    \tilde{J}_\mathrm{f} = \mathcal{O}(10^{-1}) \tilde{\chi}^{-2}_\mathrm{out,f},
\end{equation}
as derived from Eq.~\eqref{eq_scaling_mn}. 
Since we consider adiabatic evolution, the total particle number $J_\mathrm{f}$ equals $J_\mathrm{i}$, given by Eq.~\eqref{eq_start_n}. 
Consequently, we obtain
\begin{equation}
\label{eq_chiout_rate}
    \chi_\mathrm{out,f}/\chi_\mathrm{out,i} = \mathcal{O}(10).
\end{equation}
We conclude that the required change in $\chi_\mathrm{out}$ during the evolution from the initial state to the final state is 
\begin{equation}
\label{eq_deltachiout_compact}
    \Delta\tilde{\chi}_\mathrm{out} \simeq \tilde{\chi}_\mathrm{out,f}.
\end{equation}
Thus, in the region where $\tilde{\chi}_\mathrm{out,f} \gtrsim \mathcal{O}(1)$, we find that a natural initial state of a non-relativistic boson cloud can evolve into a compact boson star if the $\chi$-field undergoes a change of $\Delta\tilde{\chi}_\mathrm{out} \gtrsim \mathcal{O}(1)$.  
If $\tilde{\chi}_\mathrm{out} \gg \mathcal{O}(1)$, the required change exceeds the Planck scale significantly, making it unnatural. 
Therefore, it is necessary to consider adiabatic evolution in the region where $\tilde{\chi}_\mathrm{out,f} = \mathcal{O}(1)$. 
\par
Next, we evaluate the extent of the changes in the effective mass $m_\phi$ and the total mass $M$ during this evolution. 
The effective mass of the scalar field, $m_\phi = g \chi_\mathrm{out}$, in the initial and final states is related by
\begin{equation}
\label{eq_adiabatic_mphi}
    m_\mathrm{\phi,f} = \frac{\chi_\mathrm{out,f}}{\chi_\mathrm{out,i}} m_\mathrm{\phi,i}.
\end{equation}
Using Eq.~\eqref{eq_chiout_rate}, we deduce that $m_\phi$ increases by $\mathcal{O}(10)$ during the evolution from a non-relativistic boson cloud to a compact boson star.
From Fig.~\ref{fig_mphi0}, we find that $\hat{M}_\mathrm{f} = \mathcal{O}(10^{-1})$. 
Hence, the normalized total mass $\tilde{M}_\mathrm{f}$ is given by 
\begin{equation}
    \tilde{M}_\mathrm{f} = \mathcal{O}(10^{-1}) \tilde{\chi}^{-1}_\mathrm{out,f},
\end{equation}
as derived from Eq.~\eqref{eq_scaling_mn}. Considering the ratio between the masses in the final and initial states, we have
\begin{equation}
    \frac{M_\mathrm{f}}{M_\mathrm{i}} = \mathcal{O}(10^2) \left( \frac{\chi_\mathrm{out,f}}{\chi_\mathrm{out,i}} \right)^{-1} = \mathcal{O}(10).
\end{equation}
Thus, we conclude that the total mass of the boson star also increases by $\mathcal{O}(10)$ during this evolution.

\section{Discussion and Conclusion}
\label{sec_Discussion and Conclusion}

\begin{figure}[t]
  \centering
    \begin{subfigure}[b]{0.48\textwidth}
        \centering
        \includegraphics[width=\textwidth]{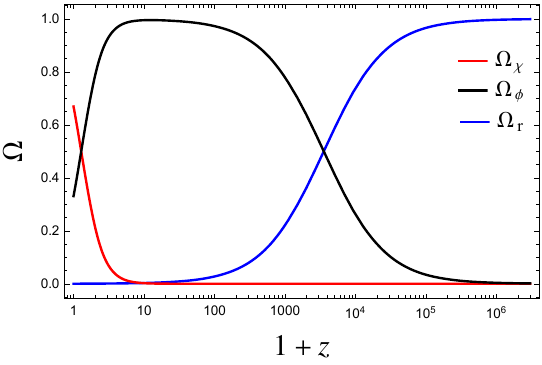}
        \caption{Case~1: the $\chi$-field acts as dark energy.}
        \label{fig_noint_de_quadr_z_omega}
    \end{subfigure}
    \hfill
    \begin{subfigure}[b]{0.48\textwidth}
        \centering
        \includegraphics[width=\textwidth]{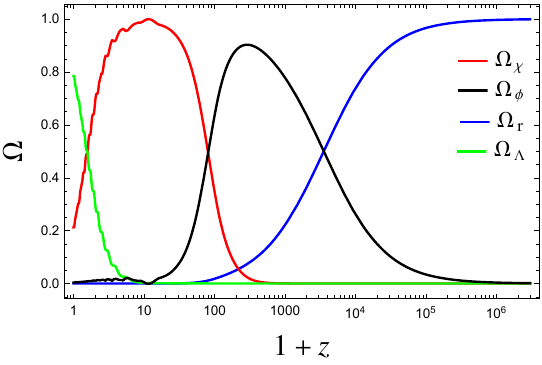}
        \caption{Case 2: the $\chi$-field acts as dark matter.}
        \label{fig_noint_dm_quadr_z_omega}
    \end{subfigure}
  \caption{
  The time evolution of each energy density parameter $\Omega$. 
  The red, black, blue, and green curves represent $\Omega_\chi(z)$, $\Omega_\phi(z)$, $\Omega_r(z)$, and $\Omega_\Lambda(z)$, respectively.
(a)The initial conditions are set at $1+z = 3 \times 10^6$, with $x = 0$, $y = 10^{-11}$, $\Omega_r = 0.99882$, and $\Omega_\Lambda = 0$. 
(b)The initial conditions are similarly chosen as $x = 0$, $y = 5 \times 10^{-9}$, $\Omega_r = 0.99882$, and $\Omega_\Lambda = 10^{-20}$.
  }
  \label{fig_noint_quadr_z_omega}
\end{figure}

Here, we evaluate the change in the $\chi$-field, $\Delta\chi_\mathrm{SR}$, that can be explained by the slow-roll within the cosmological timescale.
For a given relatively flat potential $V_\mathrm{SR}$,
\begin{equation}
\label{eq_slow_condi2}
    3H\dot{\chi}+g^2\phi^2\chi+\frac{dV_\mathrm{SR}}{d\chi}\simeq0, 
\end{equation}
with
\begin{equation}
\label{eq_slow_condi22}
    \quad H^2 = \frac{8\pi}{3m_\mathrm{pl}^2}\rho_\mathrm{tot}.
\end{equation}
follows.
If the timescale of the $\chi$-field evolution is order $H^{-1}$, the ratio of the first and second terms on the left-hand side of Eq.~\eqref{eq_slow_condi2} can be expressed as
\begin{equation}
\label{eq_condi_roll0}
\frac{g^2\phi^2}{H^2} \simeq \Omega_\mathrm{\phi} \left( \frac{m_\mathrm{pl}}{\chi_\mathrm{out}} \right)^2,
\end{equation}
where $\Omega_\phi=\rho_\phi/\rho_\mathrm{tot}$ and $\rho_\phi=g^2\chi^2\phi^2$.
Therefore, for the interaction term to be negligible in Eq.~\eqref{eq_slow_condi2},
\begin{equation}
\label{eq_condi_roll}
    \chi_\mathrm{out} \gtrsim \sqrt{\Omega_\mathrm{\phi}}m_\mathrm{pl},
\end{equation}
 must be satisfied.
 Otherwise, the $\chi$-field cannot evolve in the direction that increases the effective mass of the $\phi$-field. 
If dark matter consists entirely of the $\phi$-field, satisfying $\Omega_\phi=\Omega_\mathrm{DM}\simeq\mathcal{O}(0.1)$, this condition is consistent with the assumption that the gradient of the $\chi$-field is negligible in the initial stage, mentioned in the previous section. 
It implies that the non-gravitational interaction between the two scalar fields must be sufficiently weak.
Under this condition, Eqs.~\eqref{eq_slow_condi2}–\eqref{eq_slow_condi22} reduce to the standard slow-roll equations:
\begin{equation}
\label{eq_slow_condi}
    3H\dot{\chi} \simeq \frac{dV_\mathrm{SR}}{d\chi}, \quad H^2 \simeq \frac{3\pi}{8m_\mathrm{pl}^2}V_\mathrm{SR}.
\end{equation}
Using these equations, $\Delta\chi_\mathrm{SR}$ can be expressed as
\begin{equation}
    \Delta\chi_\mathrm{SR} \simeq H^{-1}\dot{\chi} \simeq \frac{m_\mathrm{pl}^2}{\Delta\chi_\mathrm{SR}} \frac{\Delta V_\mathrm{SR}}{V_\mathrm{SR}}.
\end{equation}
Assuming that the potential satisfies $\Delta V_\mathrm{SR}/V_\mathrm{SR} \simeq 1$, $\Delta\chi_\mathrm{SR}$ is estimated as:
\begin{equation}
\label{eq_deltachi_condi}
    \Delta\chi_\mathrm{SR} \simeq m_\mathrm{pl}.
\end{equation}
By comparing this result with Eq.~\eqref{eq_deltachiout_compact}, it can be concluded that if the $\chi$-field undergoes a slow-roll motion with $\Delta\chi_\mathrm{SR} = \mathcal{O}(m_\mathrm{pl})$, it is possible to form a compact boson star through adiabatic evolution within the cosmological timescale. 
For instance, a simple potential of the form
\begin{equation}
\label{eq_chi_potential}
    V_\mathrm{SR} = \frac{1}{2}m_\chi^2 (\chi - \chi_\mathrm{f})^2,
\end{equation}
can satisfy the assumption $\Delta V_\mathrm{SR}/V_\mathrm{SR} \simeq 1$ in the region where $\chi_\mathrm{f} = \mathcal{O}(m_\mathrm{pl})$. 

\begin{figure}[t]
  \centering
    \begin{subfigure}[b]{0.48\textwidth}
        \centering
        \includegraphics[width=\textwidth]{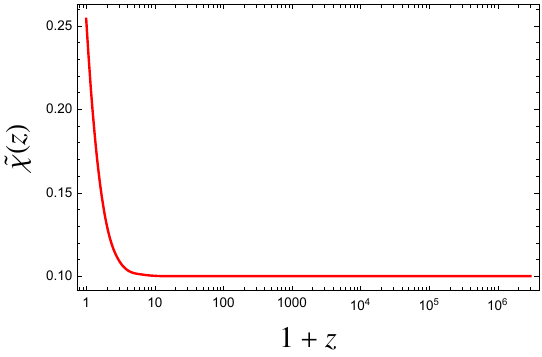}
        \caption{Case~1: the $\chi$-field acts as dark energy.}
        \label{fig_noint_de_quadr_z_chi}
    \end{subfigure}
    \hfill
    \begin{subfigure}[b]{0.48\textwidth}
        \centering
        \includegraphics[width=\textwidth]{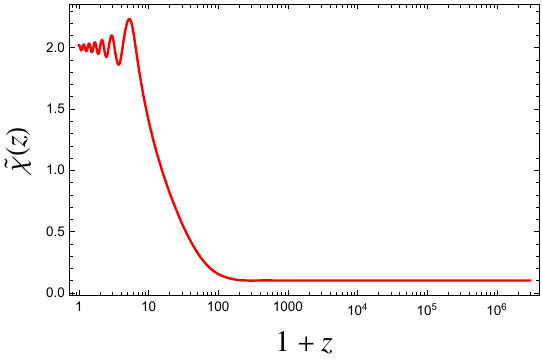}
        \caption{Case 2: the $\chi$-field acts as dark matter.}
        \label{fig_noint_dm_quadr_z_chi}
    \end{subfigure}
  \caption{
  The time evolution of the $\chi$-field. 
In both scenarios, the initial condition is set as $\tilde{\chi} = 0.1$ at $1+z = 3 \times 10^6$, and the minimum of $V_\mathrm{SR}$ is chosen to be $\tilde{\chi}_\mathrm{f} = 2$, consistent with the Adiabatic Evolution 2 (see also Fig.~\ref{fig_n-g_graph}). 
We adopt $m_\chi \simeq 10 H_0$ for Case 1 and $m_\chi \simeq 0.1 H_0$ for Case 2.
  }
  \label{fig_noint_quadr_z_chi}
\end{figure}

\par
To test the validity of the above discussion, we numerically analyze the time evolution of the cosmological background.
We adopt an axion-type potential for the $\chi$-field, as given in Eq.~\eqref{eq_chi_potential}. 
In this setup, we consider two scenarios: Case 1, where the $\chi$-field acts as dark energy at present ($m_\chi \lesssim H_0$), and Case 2, where it behaves like dark matter today, while undergoing slow-roll evolution in the past ($H_0 \lesssim m_\chi \ll m_\phi$).
The cosmological background consists of the following components: the $\chi$-field, the $\phi$-field, radiation, and the cosmological constant $\Lambda$. 
The time evolution of each component is governed by the following equations:
\begin{align}
\label{eq_eom_bg1}
 &\ddot{\chi} + 3H\dot{\chi} + \frac{dV_\mathrm{SR}}{d\chi} = 0, \\
\label{eq_eom_bg2}
 &\dot{\rho}_r + 4H\rho_r = 0, \\
\label{eq_eom_bg3}
 &\dot{\rho}_\Lambda = 0, \\
\label{eq_eom_bg4}
 &\dot{\rho}_\phi + 3H\rho_\phi = 0,
\end{align}
with the Friedmann equation
\begin{equation}
\label{eq_eom_bgh}
    H^2 = \frac{8\pi}{3m_\mathrm{pl}^2} \left( \frac{1}{2}\dot{\chi}^2 + V_\mathrm{SR} + \rho_\phi + \rho_r + \rho_\Lambda \right).
\end{equation}
Here, $\rho_r$ and $\rho_\Lambda$ denote the energy densities of radiation and the cosmological constant, respectively.
In this analysis, we neglect the interaction between the $\chi$- and $\phi$-fields for simplicity.
Following the standard approach in quintessence models (e.g.,~\cite{Tsujikawa:2013fta}), we define the following dimensionless variables:
\begin{equation}
\label{eq_slowroll_dimless}
    x = \frac{\sqrt{8\pi}\dot{\chi}}{\sqrt{6}m_\mathrm{pl}H}, \quad
    y = \frac{\sqrt{8\pi V_\mathrm{SR}}}{\sqrt{3}m_\mathrm{pl}H}, \quad
    \Omega_I = \frac{8\pi\rho_I}{3m_\mathrm{pl}^2H^2},
\end{equation}
where $I = \chi,~\phi,~r,~\Lambda$.
The energy density parameters then satisfy:
\begin{equation}
    \Omega_\chi = x^2 + y^2, \quad
    \Omega_\phi = 1 - \Omega_\chi - \Omega_r - \Omega_\Lambda.
\end{equation}
Based on this formulation, we numerically solve Eqs.~\eqref{eq_eom_bg1}--\eqref{eq_eom_bg3} under Eq.~\eqref{eq_eom_bgh}, treating $\chi$, $x$, $y$, $\Omega_r$, and $\Omega_\Lambda$ as functions of the redshift $z$.
We then evaluate the total change in the background field, denoted by $\Delta \chi_\mathrm{SR}$.

We show the time evolution of each energy density parameter $\Omega$ in Fig.~\ref{fig_noint_quadr_z_omega}. 
In both figures, the red, black, blue, and green curves represent $\Omega_\chi(z)$, $\Omega_\phi(z)$, $\Omega_r(z)$, and $\Omega_\Lambda(z)$, respectively.
Figure~\ref{fig_noint_de_quadr_z_omega} corresponds to Case 1, in which the $\chi$-field behaves as dark energy at present. 
The initial conditions are set at $1+z = 3 \times 10^6$, with $x = 0$, $y = 10^{-11}$, $\Omega_r = 0.99882$, and $\Omega_\Lambda = 0$. 
In contrast, Fig.~\ref{fig_noint_dm_quadr_z_omega} shows the evolution in Case 2, where the $\chi$-field behaves as dark matter today but was initially slow-rolling. 
The initial conditions are similarly chosen as $x = 0$, $y = 5 \times 10^{-9}$, $\Omega_r = 0.99882$, and $\Omega_\Lambda = 10^{-20}$.
In both cases, the parameters are tuned such that the present-day energy densities are consistent with observations: $\Omega_{\mathrm{DE},0} \simeq 0.7$, $\Omega_{\mathrm{m},0} \simeq 0.3$, and $\Omega_{r,0} \simeq 10^{-6}$.

We show the time evolution of the $\chi$-field in Fig.~\ref{fig_noint_quadr_z_chi}. 
In both scenarios, the initial condition is set as $\tilde{\chi} = 0.1$ at $1+z = 3 \times 10^6$, and the minimum of the background potential is chosen to be $\tilde{\chi}_\mathrm{f} = 2$, consistent with the Adiabatic Evolution 2 scenario (see also Fig.~\ref{fig_n-g_graph}). 
We adopt $m_\chi \simeq 10 H_0$ for Case 1 and $m_\chi \simeq 0.1 H_0$ for Case 2.
In Case 2 (Fig.~\ref{fig_noint_dm_quadr_z_chi}), the $\chi$-field begins to oscillate around $z \simeq 10$, which corresponds to the condition $m_\chi \simeq H(z)$ and marks the breakdown of the slow-roll approximation. 
During this period, the field evolves from $\tilde{\chi} = 0.1$ to $\tilde{\chi}_\mathrm{f} = 2$ within the cosmological timescale, indicating that compact boson stars can be formed via Adiabatic Evolution2.
In contrast, in Case 1 (Fig.~\ref{fig_noint_de_quadr_z_chi}), the $\chi$-field remains in the slow-roll regime until the present and does not enter the oscillatory phase. 
As a result, it stays around $\tilde{\chi} \simeq 0.25$, suggesting that compact boson star formation has not yet occurred in this scenario.
These results imply that, for an axion-type potential, the formation of compact boson stars requires the $\chi$-field to behave not as dark energy but rather as a light dark matter component, as in Case 2. 
A detailed analysis of the stability of the resulting boson stars after the onset of oscillations in Case 2 is left for future work. 
The effects of interactions between the scalar fields are discussed separately in Appendix~\ref{app1}.

\par
In this paper, we showed that the boson star can attain a more compact configuration via adiabatic evolution.
We found that if the $\chi$-field undergoes slow-roll motion in a direction that increases the effective mass of the $\phi$-field, the boson star becomes more compact. 
To form a compact boson star, the $\chi$-field must undergo an excursion of order of the Planck scale in general.
In this study, we assume that the gradient of the $\chi$-field is negligible in the initial stage, allowing these configurations to be well-approximated by the Schrödinger-Poisson system.
If this approximation holds, the $\phi$-field can form non-relativistic boson clouds due to gravitational scattering on the timescales given by Eq.~\eqref{eq_relaxationtime} and Eq.~\eqref{eq_relaxationtime2}.
However, in reality, it is challenging to form such boson clouds in our scenario, since $\chi_\mathrm{out,i}$ must be much smaller than $m_\mathrm{pl}$ for the gradient to be negligible.
In fact, we demonstrated that $\chi_\mathrm{out,i}$ is smaller than $m_\mathrm{pl}$ and $E_\mathrm{grad,\chi}\gtrsim E_\mathrm{grav}$ at the initial stage, even for the case of adiabatic evolution with $\chi_\mathrm{out,f}\simeq10$.
Therefore, we conclude that it is not possible to form compact boson stars from initial boson clouds prepared in the framework of the Schr\"odinger-Poisson system through adiabatic evolution.
\par
Therefore, further investigations are necessary to establish a scenario to form compact boson stars through adiabatic evolution in our universe.
In particular, we need to extend the study about the initial cloud formation by removing the assumption that the gradient of the $\chi$-field is negligible. 
However, it remains a future work to investigate how the formation of 
boson clouds is affected by interactions between the $\phi$-field and the $\chi$-field. 

On the other hand, Eq.\eqref{eq_condi_roll} implies that the initial condition for $\chi_\mathrm{out,i}$ in adiabatic evolution must satisfy
\begin{equation}
\label{eq_condi_new}
\sqrt{\Omega_\mathrm{\phi}}m_\mathrm{pl}\lesssim\chi_\mathrm{out,i}\lesssim m_\mathrm{pl}.
\end{equation}
The upper bound of this inequality stems from the constraint that $\chi_\mathrm{out} \gg m_\mathrm{pl}$ is prohibited.
Hence, if $\Omega_\phi\approx \Omega_\mathrm{DM}$, it seems difficult 
to find evolution paths that reach compact boson stars. However, 
if $\Omega_\phi\ll \Omega_\mathrm{DM}$, all the Adiabatic Evolutions 1, 2, and 3, in which $\chi_\mathrm{out,i} = \mathcal{O}(0.01 - 0.1m_\mathrm{pl})$, become relevant, satisfying the constraint~\eqref{eq_condi_new}.

\par
For simplicity, we focused on boson stars composed of a complex scalar field in this paper. 
However, oscillatons, which are boson stars formed from real scalar fields, such as axions, are also widely studied. 
Unlike boson stars composed of complex scalar fields, oscillatons are expected to be accompanied by scalar radiation. 
Due to this characteristic, it remains unclear to what extent the adiabatic approximation is applicable to oscillatons. 
Therefore, it might be interesting to investigate the possibility of forming compact boson stars through adiabatic evolution while accounting for scalar radiation, too.

\section*{Acknowledgements}
We thank Hidetoshi Omiya, Takuya Takahashi, and Daiki Saito for discussions
and useful comments.
This work was supported by JST SPRING, Grant Number JPMJSP2110(YM), JSPS KAKENHI Grant Nos. JP23H00110, 24H00963 and 24H01809(TT).
\bibliographystyle{JHEP}
\bibliography{refref}

\appendix
\section{
Numerical Analysis of the Background Evolution with Scalar-Field Interaction
}
\label{app1}

\begin{figure}[t]
  \centering
    \begin{subfigure}[b]{0.48\textwidth}
        \centering
        \includegraphics[width=\textwidth]{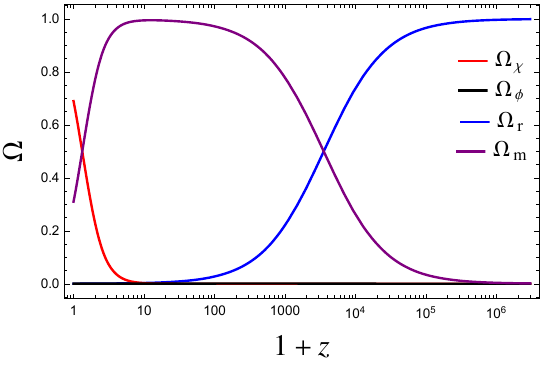}
        \caption{Case1: The $\chi$-field behaves as dark energy, interacting with the $\phi$-field.
        }
        \label{fig_int_de_quadr_z_omega}
    \end{subfigure}
    \hfill
    \begin{subfigure}[b]{0.48\textwidth}
        \centering
        \includegraphics[width=\textwidth]{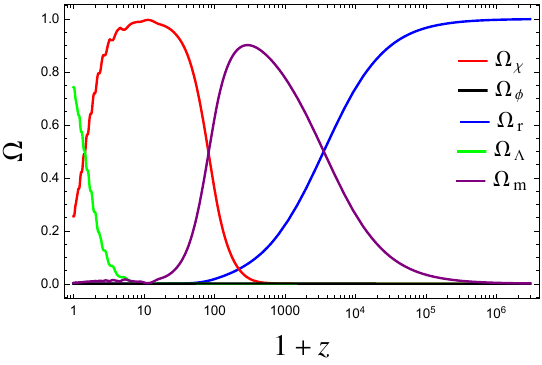}
        \caption{Case 2: the $\chi$-field acts as dark matter, interacting with the $\phi$-field.}
        \label{fig_int_dm_quadr_z_omega}
    \end{subfigure}
  \caption{
  The time evolution of each energy density parameter $\Omega$ in the presence of $\phi$–$\chi$ interaction. 
  In both figures, the red, black, blue, green, and purple curves represent $\Omega_\chi(z)$, $\Omega_\phi(z)$, $\Omega_r(z)$, $\Omega_\Lambda(z)$, and $\Omega_m(z)$, respectively.
(a)The initial conditions are set at $1+z = 3 \times 10^6$, with $x = 0$, $y = 10^{-11}$, $\Omega_r = 0.99882$, $\Omega_\Lambda = 0$, and $N_\phi=10^{-5}$. 
(b)The initial conditions are similarly chosen as $x = 0$, $y = 5\times10^{-9}$, $\Omega_r = 0.99882$, $\Omega_\Lambda = 10^{-20}$, and $N_\phi=10^{-5}$.
  }
  \label{fig_int_quadr_z_omega}
\end{figure}

\begin{figure}[t]
  \centering
    \begin{subfigure}[b]{0.48\textwidth}
        \centering
        \includegraphics[width=\textwidth]{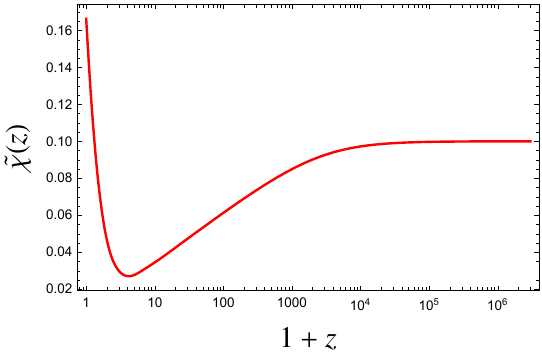}
        \caption{Case1: The $\chi$-field behaves as dark energy, interacting with the $\phi$-field.}
        \label{fig_int_de_quadr_z_chi}
    \end{subfigure}
    \hfill
    \begin{subfigure}[b]{0.48\textwidth}
        \centering
        \includegraphics[width=\textwidth]{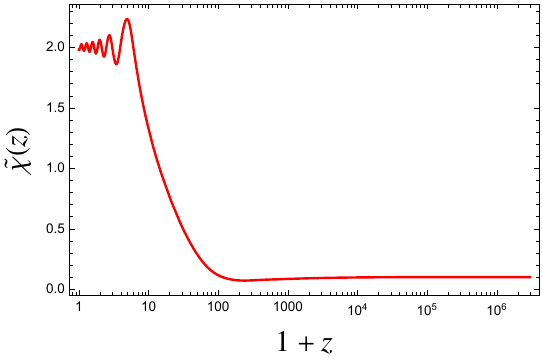}
        \caption{Case 2: the $\chi$-field acts as dark matter, interacting with the $\phi$-field.}
        \label{fig_int_dm_quadr_z_chi}
    \end{subfigure}
  \caption{
   The time evolution of the $\chi$-field in the presence of $\phi$–$\chi$ interaction. 
In both scenarios, the initial condition is set as $\tilde{\chi} = 0.1$ at $1+z = 3 \times 10^6$, and the minimum of $V_\mathrm{SR}$ is chosen to be $\tilde{\chi}_\mathrm{f} = 2$, consistent with the Adiabatic Evolution 2 (see also Fig.~\ref{fig_n-g_graph}). 
We adopt $m_\chi \simeq 10 H_0$ for Case 1 and $m_\chi \simeq 0.1 H_0$ for Case 2.
  }
  \label{fig_int_quadr_z_chi}
\end{figure}

In Sec.~\ref{sec_Discussion and Conclusion}, we evaluated the total variation of the $\chi$-field assuming no interaction with the $\phi$-field. 
However, given that $\phi^2 \propto a^{-3}$, it is not immediately clear whether the adiabatic condition in Eq.~\eqref{eq_condi_roll} remains valid throughout the evolution.
In this section, we investigate whether the necessary variation $\Delta\chi$ required for compact boson star formation can still be achieved when interactions between $\phi$ and $\chi$ are present. 
We consider the scalar potential of the form
\begin{equation}
    V(\phi,\chi) = \frac{1}{2}g^2 \phi^2 \chi^2 + V_\mathrm{SR}(\chi),
\end{equation}
where $V_\mathrm{SR}$ is taken to be of the axion type, as introduced in Eq.~\eqref{eq_chi_potential}.
The cosmological background dynamics of the scalar fields are then governed by
\begin{align}
\label{eq_ap_slow}
  &\ddot{\chi} + 3H\dot{\chi} + V_{\mathrm{SR},\chi} = -\frac{\rho_\phi}{\chi}, \\
\label{eq_ap_slow_phi}
  &\ddot{\phi} + 3H\dot{\phi} + m_\phi^2 \phi = 0,
\end{align}
where $V_{\mathrm{SR},\chi} \equiv dV_{\mathrm{SR}}/d\chi$, $\rho_\phi = g^2 \chi^2 \phi^2$, and $m_\phi = g\chi$.
\par
To estimate the strength of the interaction between the $\phi$-field and the $\chi$-field, we evaluate the ratio of the interaction term on the right-hand side of Eq.~\eqref{eq_ap_slow} to the third term on the left-hand side:
\begin{equation}
\label{eq_ap_int}
    \frac{-\rho_\phi/\chi}{V_\mathrm{SR,\chi}}\simeq\frac{\Omega_\phi}{\Omega_\chi}\frac{m_\mathrm{pl}}{\sqrt{8\pi}\chi}\lambda_\mathrm{eff}^{-1},\quad \lambda_\mathrm{eff}=\frac{-m_\mathrm{pl} V_\mathrm{SR,\chi}}{\sqrt{8\pi}V_\mathrm{SR}},
\end{equation}
where $\Omega_\phi=\rho_\phi/\rho_\mathrm{tot}$ and $\Omega_\chi\simeq V_\mathrm{SR}/\rho_\mathrm{tot}$.
If the timescale of the $\chi$-field evolution is order $H^{-1}$ and satisfies the slow-roll condition $3H\dot{\chi}\simeq -V_\mathrm{SR,\chi}$, Eq.~\eqref{eq_ap_int} can be expressed as
\begin{equation}
\frac{-\rho_\phi/\chi}{V_\mathrm{SR,\chi}} \simeq \Omega_\mathrm{\phi} \left( \frac{m_\mathrm{pl}}{\chi} \right)^2,
\end{equation}
which is consistent with Eq.~\eqref{eq_condi_roll0}.
In practice, it is difficult to ignore the interaction between scalar fields, during the evolution. 
We consider the regime where the $\chi$-field does not significantly exceed the Planck scale, i.e., $\chi/m_\mathrm{pl} \lesssim \mathcal{O}(1)$.  
We next estimate the strength of $\lambda_\mathrm{eff}$ in Eq.~\eqref{eq_ap_int}.
For example, if the $\chi$-field follows a quintessence-type potential $V_\mathrm{SR} \propto \exp(-\sqrt{8\pi}\lambda\chi/m_\mathrm{pl})$, then $\lambda_\mathrm{eff} = \lambda$. 
In this model, accelerated expansion requires $\lambda^2 < 2$~\cite{Tsujikawa:2013fta}. 
On the other hand, for an axion-type potential given in Eq.~\eqref{eq_chi_potential}, we estimate $\lambda_\mathrm{eff} \simeq m_\mathrm{pl}/(\chi_\mathrm{f}-\chi)$. 
$\chi \simeq \chi_\mathrm{i}$ at early stage, Eq.~\eqref{eq_ap_int} implies $(m_\mathrm{pl}/\chi)\lambda_\mathrm{eff}^{-1} = \mathcal{O}(10)$.
To justify neglecting the interaction term in Eq.~\eqref{eq_ap_slow}, one must ensure that $\Omega_\phi(z) \ll \Omega_\chi(z)$ throughout the evolution. 
The $\phi$-field behaves as a small fraction of cold dark matter, with $\rho_\phi$ scaling approximately as $a^{-3}$, while the $\chi$-field is sufficiently light to undergo slow-roll motion, approximately satisfying $\rho_\chi \simeq V_\mathrm{SR} \simeq \mathrm{const.}$
Consequently, unless $\rho_\phi$ is extremely small today (i.e., $\phi$ does not constitute dark matter), there must exist an epoch in the past where $\Omega_\phi(z)/\Omega_\chi(z) = \mathcal{O}(1)$, and the interaction cannot be neglected during the evolution. 
\par
Using the relation $\rho_\phi \simeq m_\phi^2 \phi^2$, Eq.~\eqref{eq_ap_slow_phi} can be rewritten as
\begin{equation}
\label{eq_ap_slow_phi2}
    \dot{\rho}_\phi = \left(-3H + \frac{\dot{\chi}}{\chi}\right)\rho_\phi.
\end{equation}
If the $\chi$-field satisfies the slow-roll condition, i.e., $\dot{\chi}/\chi \ll H$, then $\rho_\phi$ approximately redshifts as $a^{-3}$.
Furthermore, defining the number density of the background $\phi$-field as $n_\phi = \rho_\phi / m_\phi$, Eq.~\eqref{eq_ap_slow_phi2} can be rewritten as 
\begin{equation}
\label{eq_ap_slow_phi3}
    \dot{n}_\phi = -3H n_\phi,
\end{equation}
which corresponds to the expected dilution due to cosmic expansion in the adiabatic evolution.
\par
Similar to the discussion in Sec.~\ref{sec_Discussion and Conclusion}, we consider two scenarios: Case 1, where the $\chi$-field acts as dark energy, and Case 2, where it behaves like dark matter today after undergoing slow-roll evolution in the past.
Since $\Omega_\phi$ is expected to remain sufficiently small, we include an additional matter component that does not interact with the $\chi$-field, in addition to the $\chi$-field, $\phi$-field, radiation, and the cosmological constant $\Lambda$, to ensure consistency with the observed present-day values of the density parameters.
The time evolution of each component is governed by
\begin{align}
\label{eq_eom_bgap1}
 &\dot{\rho}_r + 4H\rho_r = 0,\quad \dot{\rho}_\Lambda = 0, \\
\label{eq_eom_bgap2}
 &\dot{\rho}_m + 3H\rho_m = 0,
\end{align}
in addition to Eq.~\eqref{eq_ap_slow} and Eq.~\eqref{eq_ap_slow_phi3},
with
\begin{equation}
\label{eq_eom_bgaph}
    H^2 = \frac{8\pi}{3m_\mathrm{pl}^2} \left( \frac{1}{2}\dot{\chi}^2 + V_\mathrm{SR} + \rho_\phi + \rho_r + \rho_\Lambda+\rho_m \right).
\end{equation}
We define the dimensionless variables $x$, $y$, and $\Omega_I$ as in Eq.~\eqref{eq_slowroll_dimless}, and introduce
\begin{equation}
    N_\phi=\frac{8\pi n_\phi}{3m_\mathrm{pl}^2H^2}.
\end{equation}
The energy density parameters then satisfy:
\begin{equation}
    \Omega_\chi = x^2 + y^2, \quad
    \Omega_\phi =m_\phi N_\phi ,\quad
    \Omega_m = 1 - \Omega_\chi - \Omega_r - \Omega_\Lambda-\Omega_\phi.
\end{equation}
Using the dimensionless variables defined above, Eqs.~\eqref{eq_ap_slow},~\eqref{eq_ap_slow_phi3} and~\eqref{eq_eom_bgap1} can be rewritten as
\begin{align}
    \label{eq_eom_bgapx}
 &x'=-3x-\frac{\sqrt{6}}{2}\lambda_\mathrm{eff}y^2-x\frac{\dot{H}}{H^2}-\frac{\sqrt{6}}{2}N_\phi, \\
  \label{eq_eom_bgapy}
 &y'=y\left(\frac{\sqrt{6}}{2}\lambda_\mathrm{eff}x-\frac{\dot{H}}{H^2}\right), \\
 \label{eq_eom_bgapr}
 &\Omega_r'=-\Omega_r\left(4+2\frac{\dot{H}}{H^2}\right), \\
 \label{eq_eom_bgaplambda}
 &\Omega_\Lambda'=-2\Omega_\Lambda\frac{\dot{H}}{H^2}, \\
 \label{eq_eom_bgapnphi}
 &N_\phi'=-N_\phi\left(3+2\frac{\dot{H}}{H^2}\right), \\
\end{align}
where $f'=df/d\ln a$ and
\begin{equation}
\label{eq_eom_bgaph2}
    \frac{\dot{H}}{H^2}=-3x^2-\frac{3}{2}\Omega_m-2\Omega_r-\frac{3}{2}\Omega_\phi.
\end{equation}
By setting $N_\phi = 0$ and replacing $m$ with $\phi$, Eqs.~\eqref{eq_eom_bgapx}--\eqref{eq_eom_bgaph2} reduce to the set of equations used in Sec.~\ref{sec_Discussion and Conclusion}, where the interaction was neglected.
To examine the impact of the interaction term, we numerically solve Eqs.~\eqref{eq_eom_bgapx}--\eqref{eq_eom_bgaph2} under suitable initial conditions.
\begin{figure}[t]
\centering
\includegraphics[width=0.6\linewidth]{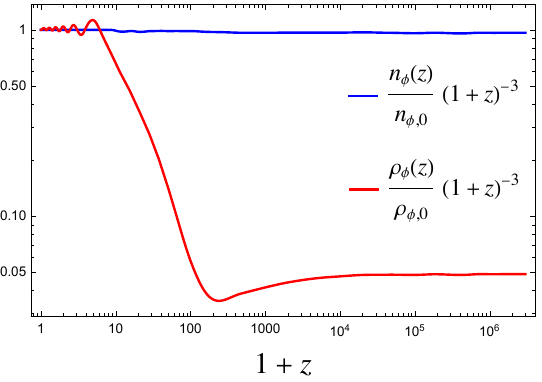}
\caption{
The time evolution of $n_\phi$ and $\rho_\phi$ for Case~2.
The blue and red curves represent the normalized quantities $n_\phi / [n_{\phi,0}(1+z)^3]$ and $\rho_\phi / [\rho_{\phi,0}(1+z)^3]$, respectively.
$n_{\phi,0}$ and $\rho_{\phi,0}$ denote the respective present-day values.
}
\label{fig_int_rho_z_omega}
\end{figure}

\par
We show the time evolution of each energy density parameter $\Omega$ in Fig.~\ref{fig_int_quadr_z_omega}. 
In both figures, the red, black, blue, green, and purple curves represent $\Omega_\chi(z)$, $\Omega_\phi(z)$, $\Omega_r(z)$, $\Omega_\Lambda(z)$, and $\Omega_m(z)$, respectively.
Figure~\ref{fig_int_de_quadr_z_omega} corresponds to Case 1. 
The initial conditions are set at $1+z = 3 \times 10^6$, with $x = 0$, $y = 10^{-11}$, $\Omega_r = 0.99882$, $\Omega_\Lambda = 0$, and $N_\phi=10^{-5}$. 
In contrast, Fig.~\ref{fig_int_dm_quadr_z_omega} shows the evolution in Case 2. 
The initial conditions are similarly chosen as $x = 0$, $y = 5\times10^{-9}$, $\Omega_r = 0.99882$, $\Omega_\Lambda = 10^{-20}$, and $N_\phi=10^{-5}$.
In both cases, the parameters are tuned such that the present-day energy densities are consistent with observations: $\Omega_{\mathrm{DE},0} \simeq 0.7$, $\Omega_{\mathrm{m},0} \simeq 0.3$, and $\Omega_{r,0} \simeq 10^{-6}$.
We show the time evolution of the $\chi$-field in Fig.~\ref{fig_int_quadr_z_chi}. 
To ensure consistency with the Adiabatic Evolution2, we fix the potential minimum at $\tilde{\chi}_\mathrm{f} = 2$ and set the initial condition as $\tilde{\chi} = 0.1$ at $1 + z = 3 \times 10^6$ for both cases. 
We adopt $m_\chi \simeq 10 H_0$ for Case 1 (Fig.~\ref{fig_noint_de_quadr_z_chi}) and $m_\chi \simeq 0.1 H_0$ for Case 2 (Fig.~\ref{fig_noint_dm_quadr_z_chi}).
The initial conditions and model parameters used in Fig.~\ref{fig_int_quadr_z_omega} and Fig.~\ref{fig_int_quadr_z_chi} are identical to those in Sec.~\ref{sec_Discussion and Conclusion}, except for the inclusion of $N_\phi$. 
As a result, the time evolution of the density parameters $\Omega_I$ and the $\chi$-field  closely follows the behavior observed in the non-interacting case. 
This similarity suggests that the qualitative discussions in Sec.~\ref{sec_Discussion and Conclusion} approximately remain valid even when interactions are included.
However, as shown in Fig.~\ref{fig_int_quadr_z_chi}, there exists a period during which the value of $\chi$ decreases—for instance, $4 \lesssim 1+z \lesssim 10^4$ in Fig.~\ref{fig_noint_de_quadr_z_chi}.
This deviation from the non-interacting case arises due to the interaction term $g^2 \phi^2 \chi^2$, which effectively pushes the field toward smaller values.
We therefore evaluate the quantitative impact of this interaction on the evolution of $\rho_\phi$ and the reduction in $\chi$.

\par
We show the time evolution of $n_\phi$ and $\rho_\phi$ for Case~2 in Fig.~\ref{fig_int_rho_z_omega}.
The blue curve shows $n_\phi / [n_{\phi,0}(1+z)^3]$, and the red curve indicates $\rho_\phi / [\rho_{\phi,0}(1+z)^3]$, where $n_{\phi,0}$ and $\rho_{\phi,0}$ denote the respective present-day values.
As seen in the figure, $n_\phi$ strictly follows the expected scaling $n_\phi \propto a^{-3}$ throughout cosmic history, consistent with Eq.~\eqref{eq_ap_slow_phi3}. 
In contrast, $\rho_\phi$ approximately evolves as $\rho_\phi \propto a^{-3}$ during the slow-roll and oscillatory phases of the $\chi$-field (see Fig.~\ref{fig_int_dm_quadr_z_chi}).
However, during the transition between these regimes, $\rho_\phi$ exhibits an enhancement by a factor of $\mathcal{O}(10)$.
This behavior demonstrates that the evolution of $\rho_\phi$ is directly affected by the dynamics of the $\chi$-field through their mutual interaction.
\par
\begin{figure}[t]
  \centering
    \begin{subfigure}[b]{0.48\textwidth}
        \centering
        \includegraphics[width=\textwidth]{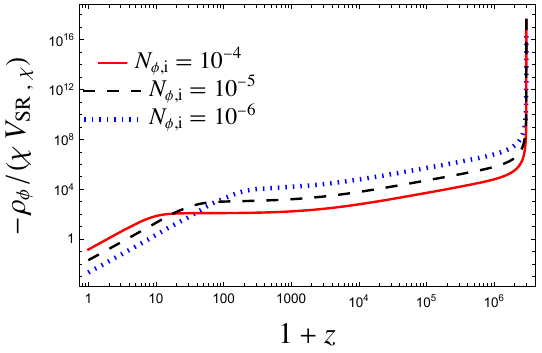}
        \protect\phantomcaption
        \label{fig_int_de_int_z_int}
        \raisebox{0.5\height}{\hspace{1.1cm} \large \textbf{(a)}}
    \end{subfigure}
    \hfill
    \begin{subfigure}[b]{0.48\textwidth}
        \centering
        \includegraphics[width=\textwidth]{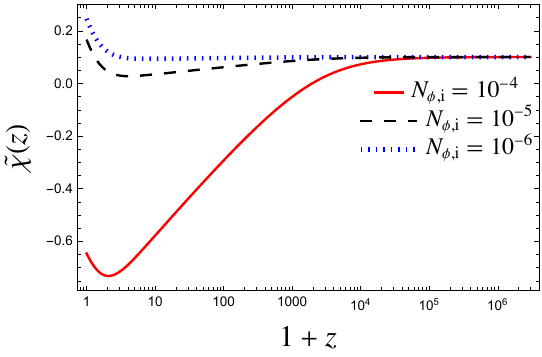}
        \protect\phantomcaption
        \label{fig_int_de_int_z_chi}
        \raisebox{0.5\height}{\hspace{1.1cm} \large \textbf{(b)}}
    \end{subfigure}
  \caption{
  The time evolution of both $-\rho_\phi / (\chi V_{\mathrm{SR},\chi})$ [see Eq.~\eqref{eq_ap_int}] and the $\chi$-field for various initial values of $N_\phi$ in Case~1.
In each panel, the red, black, and blue curves correspond to $N_{\phi,\mathrm{i}} = 10^{-4},\ 10^{-5},\ 10^{-6}$, respectively, where $N_{\phi,\mathrm{i}}$ denotes the initial value of $N_\phi$.
Except for $N_\phi$, all other initial conditions are identical to those in Fig.~\ref{fig_int_de_quadr_z_omega}.
  }
  \label{fig_int_de_int_z.pdf}
\end{figure}

We next investigate how the interaction term affects the dynamics of the $\chi$-field.
We show the time evolution of both $-\rho_\phi / (\chi V_{\mathrm{SR},\chi})$ [see Eq.~\eqref{eq_ap_int}] and the $\chi$-field for various initial values of $N_\phi$ in Case~1 in Fig.~\ref{fig_int_de_int_z.pdf}.
In each panel, the red, black, and blue curves correspond to $N_{\phi,\mathrm{i}} = 10^{-4},\ 10^{-5},\ 10^{-6}$, respectively, where $N_{\phi,\mathrm{i}}$ denotes the initial value of $N_\phi$.
Except for $N_\phi$, all other initial conditions are identical to those in Fig.~\ref{fig_int_de_quadr_z_omega}.
From Fig.~\ref{fig_int_de_int_z_int}, we observe that the redshift below which the interaction becomes negligible (i.e., where $-\rho_\phi / (\chi V_{\mathrm{SR},\chi}) \lesssim 1$) shifts to lower values as $N_{\phi,\mathrm{i}}$ increases.
Additionally, Fig.~\ref{fig_int_de_int_z_chi} shows that larger initial values of $N_\phi$ lead to a more significant decrease in the $\chi$-field over time.
These results indicate that a larger $N_\phi$ enhances the impact of the interaction, driving the evolution of $\chi$ toward smaller values.
In particular, for $N_{\phi,\mathrm{i}} = 10^{-4}$, the $\chi$-field evolves to negative values.
Even for $N_{\phi,\mathrm{i}} = 10^{-5}$, $\Omega_{\phi,0}$ is $\mathcal{O}(10^{-4})$, implying that the $\phi$-field must contribute negligibly to the total dark matter abundance in order for this scenario to permit the formation of compact boson stars.
\par
In summary, our analysis shows that even in the presence of interactions, the cosmological evolution of the $\chi$-field can be sufficient to reach the conditions required for the formation of compact boson stars.
However, this scenario requires the $\chi$-field to behave as ultra-light dark matter, and the contribution of the $\phi$-field to the total dark matter abundance must remain subdominant.

\section{Definitions of the Radius and Compactness of Boson Stars}
\label{sec_Definitions of the Radius and Compactness of Boson Stars}
\begin{figure}[t]
  \centering
    \begin{subfigure}[b]{0.48\textwidth}
        \centering
        \includegraphics[width=\textwidth]{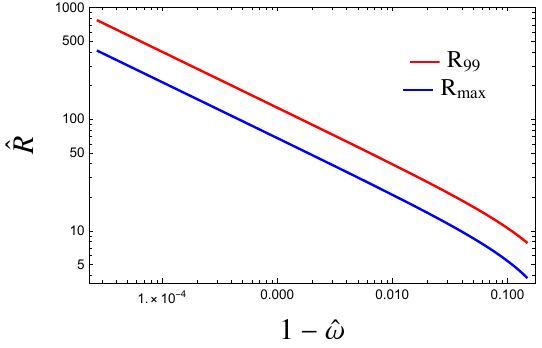}
        \protect\phantomcaption
        \label{fig_rplotmini}
        \raisebox{0.5\height}{\hspace{1.1cm} \large \textbf{(a)}}
    \end{subfigure}
    \hfill
    \begin{subfigure}[b]{0.48\textwidth}
        \centering
        \includegraphics[width=\textwidth]{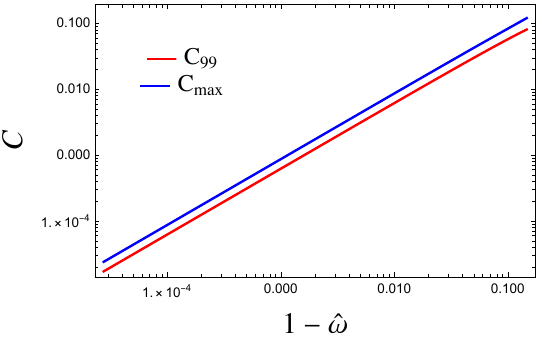}
        \protect\phantomcaption
        \label{fig_cplotmini}
        \raisebox{0.5\height}{\hspace{1.1cm} \large \textbf{(b)}}
    \end{subfigure}
  \caption{The differences between the two radius definitions and the corresponding compactness values for
        the series of the mini-boson star. The horizontal axis represents $1 - \hat{\omega}$. In the left panel, the red
        and blue curves correspond to $\hat{R}_{99}$ and $\hat{R}_{\mathrm{max}}$, respectively. In the right panel, the red and blue
        curves represent $C_{99}$ and $C_{\mathrm{max}}$, respectively.
  }
  \label{fig_plotmini}
\end{figure}

\begin{figure}[t]
  \centering
    \begin{subfigure}[b]{0.48\textwidth}
        \centering
        \includegraphics[width=\textwidth]{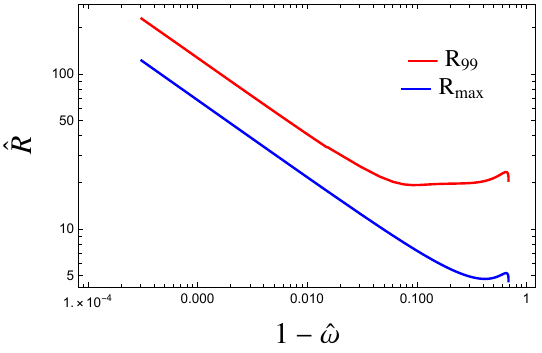}
        \protect\phantomcaption
        \label{fig_rplot05}
        \raisebox{0.5\height}{\hspace{1.1cm} \large \textbf{(a)}}
    \end{subfigure}
    \hfill
    \begin{subfigure}[b]{0.48\textwidth}
        \centering
        \includegraphics[width=\textwidth]{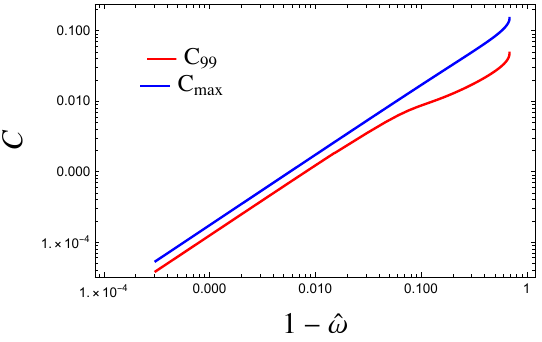}
        \protect\phantomcaption
        \label{fig_cplot05}
        \raisebox{0.5\height}{\hspace{1.1cm} \large \textbf{(b)}}
    \end{subfigure}
  \caption{
  The differences between the two radius definitions and the corresponding compactness values for
        the series of the boson star with $\tilde{\chi}_{\mathrm{out}} = 0.5$. The horizontal axis represents $1 - \hat{\omega}$. In the left
        panel, the red and blue curves correspond to $\hat{R}_{99}$ and $\hat{R}_{\mathrm{max}}$, respectively. In the right panel, the
        red and blue curves represent $C_{99}$ and $C_{\mathrm{max}}$, respectively.
  }
  \label{fig_plot05}
\end{figure}

Since boson stars composed of scalar fields lack a well-defined surface, the definition of their radius $R$ involves a degree of arbitrariness. Consequently, the compactness $C$, often used to characterize compact objects and defined as  
\begin{equation}
    C = \frac{M}{R m_\mathrm{pl}^2},
\end{equation}  
depends on the chosen definition of $R$. 
Different studies adopt various definitions of the radius of boson stars.
A commonly used approach defines $R$ as the radius $r = R_{99}$ that encloses 99\% of the total mass $M$, where the mass function~\eqref{eq_mass_def} satisfies
\begin{equation}
\label{eq_radius_def}
M(R_{99}) = \frac{99}{100} M.
\end{equation}
Thus, the compactness of boson stars is often expressed as $C_{99} = M/(R_{99} m_\mathrm{pl}^2)$
\par
In this paper, however, we define the compactness and radius differently, as described in Sec.~\ref{sec_Analysis of Stationary Configurations}.
We define the compactness of the boson star as the maximum value of Eq.~\eqref{eq_compactness_def}, denoted as $C_\mathrm{max}$, and the radius as the corresponding value of $r$, denoted as $R_\mathrm{max}$.
For standard spherically symmetric stars, which are regular at the origin and described by the Schwarzschild metric outside the star, the $rr$-component of the metric, $G^2(r)$, increases monotonically inside the star and decreases outside it, reaching its maximum at the star’s surface $r = R$.
In such cases, the compactness $C = M/(R m_\mathrm{pl}^2)$ coincides with $C_\mathrm{max}$.
To examine the differences between these definitions of the radius and compactness of the boson star, we compare the two radii and compactness for two specific cases: the mini-boson star and a boson star characterized by $\tilde{\chi}_\mathrm{out} = 0.5$, where the gradient of the $\chi$-field cannot be neglected.
\par  
In Fig.~\ref{fig_plotmini} and Fig.~\ref{fig_plot05}, we show the differences between the two radius definitions and the corresponding compactness values for the series of the mini-boson star and the boson star with $\tilde{\chi}_\mathrm{out} = 0.5$. 
The horizontal axis represents $1 - \hat{\omega}$.
In the left panels of each figure, the red and blue curves correspond to $\hat{R}_{99}$ and $\hat{R}_\mathrm{max}$, respectively.
In the right panels, the red and blue curves represent $C_{99}$ and $C_\mathrm{max}$, respectively.
From Fig.~\ref{fig_plotmini}, we observe that, for the mini-boson star, the relationship $R_{99} \simeq 2R_\mathrm{max}$ holds across all values of $\omega$.
Consequently, the two corresponding compactness satisfy $C_{99} \simeq 2C_\mathrm{max}$.  
From Fig.~\ref{fig_plot05}, we observe that in regions where $\omega$ is sufficiently large, the boson star with $\hat{\chi}=0.5$ maintains the same relationships between the two radii and compactness as those found for the mini-boson star. 
However, in regions where $\omega$ is sufficiently small, $R_{99}$ becomes significantly larger than $R_\mathrm{max}$. 
As a result, $C_{99}$ is underestimated compared to $C_\mathrm{max}$ in this regime.
\par  
Based on the above results, we find that for the mini-boson star, the radius and compactness can be evaluated at the same order of magnitude, regardless of the chosen definition.  
In contrast, for boson stars where the gradient of the $\chi$-field is significant, the radius and compactness vary significantly depending on the definition used, particularly in relativistic regions.
As shown in Fig.~\ref{fig_chioutpro}, the smaller the value of $\chi_\mathrm{out}$, the broader the profile of $\chi(r)$ compared to that of $\phi(r)$.  
On the other hand, the radial position $r = R_\mathrm{max}$, corresponding to the maximum value of $G^2(r)$, is comparable to the width of $\phi(r)$. 
Additionally, as shown in Fig.~\ref{fig_ploerate}, the smaller the value of $\chi_\mathrm{out}$, the larger the contribution of the gradient energy of the $\chi$-field to the total mass of the boson star.  
Therefore, as $\chi_\mathrm{out}$ decreases, the radius $R_{99}$ defined by Eq.~\eqref{eq_radius_def} becomes significantly larger than $R_\mathrm{max}$. 
Consequently, the compactness $C_{99}$ is underestimated compared to $C_\mathrm{max}$.
Thus, as $\chi_\mathrm{out}$ decreases, the evaluation using $R_{99}$ and $C_{99}$ becomes less reliable.  
For these reasons, we adopt $R_\mathrm{max}$ and $C_\mathrm{max}$ as the definitions of the radius and compactness, respectively, in this study.

\end{document}